\documentclass{emulateapj-rtx4}

\usepackage{lscape}
\usepackage{amsmath}
\usepackage{natbib}
\usepackage{url}
\newcommand{\ts}{\thinspace}
\newcommand{\simless}{\mathbin{\lower 3pt\hbox
     {$\rlap{\raise 5pt\hbox{$\char'074$}}\mathchar"7218$}}}
\newcommand{\simgreat}{\mathbin{\lower 3pt\hbox
     {$\rlap{\raise 5pt\hbox{$\char'076$}}\mathchar"7218$}}}
\newcommand{\msun}{\ts M$_\odot$}
\newcommand{\lsun}{\ts L$_\odot$}
\newcommand{\teff}{T$_{\rm eff}$}
\newcommand{\vsini}{$v$sin$i$}

\shortauthors{Pecaut et~al.}
\shorttitle{F-Type Members of Sco-Cen}

\begin{document}
\title{A Revised Age for Upper Scorpius and The Star-Formation History 
Among the F-Type Members of the Scorpius-Centaurus OB Association}
\author{Mark J. Pecaut\altaffilmark{1}, Eric E. Mamajek\altaffilmark{1,3}, Eric J. Bubar\altaffilmark{1,2}}
\altaffiltext{1}{University of Rochester, Department of Physics and Astronomy, Rochester, NY, 14627-0171, USA}
\altaffiltext{2}{Marymount University, Department of Biology and Physical Sciences, Arlington, VA 22207-4299,USA}
\altaffiltext{3}{Current address: Cerro Tololo Inter-American Observatory, Casilla 603, La Serena, Chile}

\begin{abstract}
We present an analysis of the ages and star-formation history of the F-type stars 
in the Upper Scorpius (US), Upper Centaurus-Lupus (UCL) and Lower Centaurus-Crux (LCC)
subgroups of Scorpius-Centaurus (Sco-Cen), the nearest OB association.
Our parent sample is the kinematically-selected
{\it Hipparcos} sample of \citet{1999AJ....117..354D}, restricted to the
138 F-type members.  We have obtained classification-resolution optical
spectra and have also determined the spectroscopic accretion disk fraction.  With 
{\it Hipparcos} and {\it 2MASS} photometry, we estimate the reddening and extinction for each star
and place the candidate members on a theoretical H-R diagram.  For each subgroup 
we construct empirical isochrones and compare to published evolutionary
tracks.  We find that 1) our empirical isochrones are consistent with the
previously published age-rank of the Sco-Cen subgroups, 2) subgroups LCC and UCL appear
to reach the main sequence turn-on at spectral types $\sim$F4 and
$\sim$F2, respectively.  An analysis of the A-type stars shows US reaching the main 
sequence at about spectral type $\sim$A3. 3) The median ages for the pre-main sequence members
of UCL and LCC are 16~Myr and 17~Myr, respectively, in agreement with previous
studies, however we find that 4) Upper Sco is much older than previously thought.
The luminosities of the F-type stars in US are typically a factor of $\sim$2.5 less luminous 
than predicted for a 5~Myr old population for four sets of evolutionary tracks. 
We re-examine the evolutionary state and isochronal ages for
the B-, A-, and G-type Upper Sco members, as well as the evolved M supergiant Antares,
and estimate a revised mean age for Upper Sco of 11$\pm$1$\pm$2~Myr (statistical, systematic). 
Using radial velocities and {\it Hipparcos} parallaxes we calculate a lower limit on the 
kinematic expansion age for Upper Sco of $>$10.5~Myr (99\% confidence).  However, the 
data are statistically consistent with no expansion.  We reevaluate the inferred masses 
for the known substellar companions in Upper Sco using the revised age and find the
inferred masses are typically $\sim$20--70\% higher than the original estimates which had
assumed a much younger age; specifically, we estimate the mass of 1RXS~J1609-2105b to
be 14$^{+2}_{-3}$~M$_\mathrm{Jup}$, suggesting that it is a brown dwarf rather than a
planet.
Finally, we find the fraction of F-type stars 
exhibiting H$\alpha$ emission and/or a K-band excess consistent with accretion to be 
0/17 ($<19$\%; 95\% C.L.) in US at $\sim$11~Myr, while UCL has 1/41 
($2^{+5}_{-1}$\%; 68\% C.L.) accretors and LCC has 1/50 ($2^{+4}_{-1}$\%; 68\% C.L.) 
accretors at $\sim$16~Myr and $\sim$17~Myr, respectively.
\end{abstract}
\maketitle

\section{Introduction and Background}
Embedded clusters hosting massive stars dominate the star formation of 
the galaxy \citep{2003ARA&A..41...57L}, and OB associations appear to be the
unbound ``fossils'' of dissolved embedded clusters.  Subgroups within OB associations 
are thought to be an approximately ``coeval'' stellar population \citep[e.g.,][]{2007prpl.conf..345B}, 
sharing a common age, chemical abundance and velocity.
Ages of stellar populations are a critical key to determining the evolutionary 
timescales of various stages of circumstellar disk evolution and, together with the
accretion disk fraction, constrain planet formation timescales 
\citep[e.g.,][]{2011ApJ...738..122C,2009AIPC.1158....3M}.  In addition, 
the estimated masses for very low-mass companions to members of the stellar population 
depend critically on the assumed age \cite[e.g.,][]{2008ApJ...689L.153L,2011ApJ...726..113I}.  
It is therefore important to estimate a consistent age for the stellar
population so it can be used with confidence for such calculations, and to periodically
test this derived age with constantly improving observational data against 
modern theoretical evolutionary models.

In this paper, we explore the F-type ($\sim$1.2--1.7 \msun\, for ages 
$\sim$10-15 Myr) stars of Scorpius-Centaurus (Sco-Cen), the
nearest OB association \citep{2008hsf2.book..235P}.  Sco-Cen consists of three subgroups:
Upper Scorpius (US), Upper Centaurus-Lupus (UCL), and Lower Centaurus-Crux
(LCC), with mean distances of 145pc, 140pc and 118pc, respectively
\citep{1999AJ....117..354D}.  As the nearest site of recent massive star
formation, the stellar membership of Sco-Cen offers important samples of young stars
for understanding circumstellar disk evolution across the mass spectrum.  

Ages for the Sco-Cen subgroups have been determined primarily from the main sequence 
turn-off and the low-mass pre-main sequence population combining H-R diagram positions
with theoretical evolutionary tracks.  For the high-mass stars, this has been performed 
most recently for UCL and LCC by \cite{2002AJ....124.1670M}, obtaining ages $\sim$17~Myr 
and $\sim$16~Myr, respectively.  The most recent published age-dating of the massive 
stars of US was performed by \cite{1989A&A...216...44D}, obtaining an age of 
$\sim$5~Myr using the evolutionary tracks of \cite{1981A&A...102..401M} and 
corroborated by \citet{2002AJ....124..404P}.

Our goal in this paper is to explore the star-formation history and accretion disk fraction
of the F-type members of three subgroups of
Scorpius-Centaurus: Upper Scorpius (US), Upper Centaurus-Lupus (UCL) and Lower
Centaurus-Crux (LCC).  Specifically, we have employed the kinematically-selected {\it Hipparcos} 
sample from \cite{1999AJ....117..354D} and have
1) placed them on a theoretical H-R diagram, 
2) used published evolutionary tracks to obtain ages and masses, 
3) compared the star-formation history of the F-type stars with previous results 
\citep{2002AJ....124.1670M,2002AJ....124..404P}, and
4) estimated the spectroscopic accretion disk fraction to constrain the disk dispersal 
timescale for intermediate mass stars.

\section{Sample Selection}
Our sample of Sco-Cen F-stars was taken from the kinematic analysis of
\cite{1999AJ....117..354D} which used both a refurbished convergent point method
\citep{1999MNRAS.306..381D} and the ``spaghetti method'' \citep{1999MNRAS.306..394H} 
to identify members based on {\it Hipparcos} positions, parallax and proper motions.  
The kinematic selection methods are described in detail in \citet{1999AJ....117..354D}, 
but briefly, the selection applied the convergent point method and spaghetti method
independently to early-type stars to create a list of ``secure'' members of the OB 
association.  A kinematic solution is obtained using both methods and the results of 
this are applied to a broader collection of stars in the {\it Hipparcos} catalog, 
with appropriate position, proper motion and distance limits for the particular
OB association.  Stars selected by both methods are included in their
membership lists, and they also provide an estimate of the overall
number of interlopers present in the membership list for each subgroup.  
For our sample (F-type candidate members only) the number of expected interlopers is
$\sim$5 in US, $\sim$15 in UCL and $\sim$9 in LCC.  Further critical examination
of the F-type samples was conducted by \citet{2011ApJ...738..122C}.

\section{Observations and Data}
Low-resolution blue ($\sim$3700\text{\AA}--5200\text{\AA}) and red 
($\sim$5600\text{\AA}--6900\text{\AA}) optical spectra were obtained from the 
SMARTS 1.5m telescope at Cerro Tololo Inter-American Observatory (CTIO).  Observations 
were made in queue mode with the RC spectrograph between February 2009 and December 2009.  
The blue spectra were obtained with a 600 grooves~mm$^{-1}$ grating (designated 26/Ia) 
blazed at 4450\text{\AA} and no filter, while the red spectra were taken with a 
831 grooves~mm$^{-1}$ grating (47/Ib) blazed at 7100\text{\AA} and GG495 filter.
Both setups use a slit width of 110.5$\mu$m.
One comparison lamp of HeAr and Neon for blue and red, respectively, was 
taken immediately before three consecutive exposures of each target.
The blue spectra has a resolution of 4.3~\text{\AA} or R $\sim$1100 
at H$\beta$ and the red spectra has a resolution of 3.1~\text{\AA} or 
R $\sim$2100 at H$\alpha$.

The data were reduced using Fred Walter's SMARTS RC Spectrograph IDL pipeline\footnote{
http://www.astro.sunysb.edu/fwalter/SMARTS/smarts\_15msched.html\#RCpipeline}.
The three object images are median combined, bias-trimmed, overscan- and bias-subtracted,  
then flat-fielded and wavelength-calibrated. Finally, we normalize the spectra to 
the continuum with a 6th order spline in preparation for spectral classification.

We have adopted V magnitudes and B-V and V-I$_C$ colors from the {\it Hipparcos} \citep{1997ESASP1200.....P} 
catalog and JHK$_S$ photometry from {\it 2MASS} \citep{2006AJ....131.1163S}.  We have taken proper
motion data, used to calculate a kinematically improved parallax (more on this in section 
\ref{sct:dist}) from both the revised {\it Hipparcos} data \citep{2007A&A...474..653V} and the 
Tycho-2 \citep{2000A&A...355L..27H} catalogs.  We used revised {\it Hipparcos} data 
if the fit obtained is that for a single star (i.e., a five-parameter solution with no 
peculiarities).  Otherwise we use the long-baseline proper motions from the Tycho-2 catalog.  The 
motivation behind this choice is to use the most precise and accurate 
proper motion data to calculate the most accurate kinematic parallax possible.  
The presence of a companion or other peculiarities
present in the astrometric solution could significantly influence the detected proper motions 
if a three-year baseline is used, as with the {\it Hipparcos} data, whereas the longer baseline 
proper motions from the Tycho-2 catalog should be less affected.  Our input data is listed in 
Table~\ref{tbl:inputparams}.

\section{Analysis}

\subsection{Spectral Classification\label{sct:spt}}
The optical spectra were visually classified using the primary temperature
classification spectral features for F-type stars: the strength and profile of
the Balmer lines, followed by several metal lines including the G-band at
$\sim$4310\text{\AA} \citep{2009ssc..book.....G}.  Pre-main sequence stars may 
exhibit enhanced chromospheric activity, which will weaken the strength of the 
Balmer absorption lines or they may be seen in emission.  In this case a correct classification 
can still be obtained using the wings of the hydrogen line profiles (Gray, private communication 2010).  
In addition to the hydrogen lines and the strength and shape of the G-band, 
we used the strength of the temperature-sensitive Ca~I $\lambda$4226 and 
Fe~I $\lambda$4383 lines to confirm the classifications, although in this
temperature region we found them less useful than the G-band.

Stellar spectral classification is greatly facilitated by the ability to
visually overlay the object on spectral standards obtained with the same spectral
resolution.  To that end, we developed a visual classification tool ({\it sptool})
which presents the object spectrum against two standard star spectra and
allows the user to change the comparison standards with a single key stroke.  The tool
allows the classifier to easily move in temperature and luminosity subtype space 
by using the arrow keys on the keyboard.  The tool, written in Python, is freely available
online\footnote{http://www.pas.rochester.edu/$\sim$mpecaut/sptool/} and only
requires a dense grid of spectral standard stars.

To obtain accurate spectral classifications for our program stars, we obtained a
grid of optical spectra of F-type MK spectral standard stars with the same 
telescope and setup as our program stars.  Our choices for 
standards are listed in Table~\ref{tbl:spstands}.  Although there are
many choices for spectral standards, ours were chosen based on a careful
consideration of the consistency of previous classifications and 
accessibility from the SMARTS 1.5m telescope at CTIO. 

Unfortunately, A8V and F4V spectral standards do not appear in the
literature (e.g. \citet{1973ARA&A..11...29M,2009ssc..book.....G}), and
the only A9V standard that we could find in the literature (44~Cet =
HR~401; \citet{1989ApJS...69..301G}) had several discrepant published
spectral types. For subtypes A8V and A9V we adopted a star that had been
assigned that spectral class by at least one expert classifier and had
colors representative of other stars of the same classification. We
adopted the Hyades member HD~27561 as a F4V standard. The star was
originally classified as F4V by \citet{1965ApJ...141..177M}, and its Hipparcos B-V
color (0.41) is intermediate in color between the F3V and F5V
standards classified by Morgan (and retained as standards by 
\citet{1973AJ.....78..386M}, \citet{1978rmsa.book.....M}, and \citet{1989AJ.....98.1049G}) 
-- HD~26015 (F3V standard, B-V=0.40) and HD~27524 (F5V standard, B-V=0.434) -- both of
which are also Hyads. We have confirmed that its spectrum is
intermediate between the F3V and F5V standards.  For the A8V standard,
we adopt the star HD~158352 (HR~6507), which was classified as A8V by
\citet{1969AJ.....74..375C}, \citet{1995ApJS...99..135A}, and as A8Vp by \citet{2001A&A...378..116M}. 
For the A9V standard, we adopted the Praesepe member HD~73450 
(BS~Cnc), based on its A9V classifications by \citet{1956PASP...68..318B},
\citet{1986PASP...98..307A}, and \citet{2002AJ....124..989G}. We verified that the
adopted A8V and A9V ``standards'' had optical spectra that were morphologically 
intermediate between the A7V and F0V MK standards.  

With our visual classification tool
and a complete grid of dwarfs, we iteratively classified the 138 program stars at least
four times before settling on a final spectral type.  We conservatively
estimate our classifications to be correct to within one subtype.

\begin{figure}
\begin{center}
\includegraphics[scale=0.45]{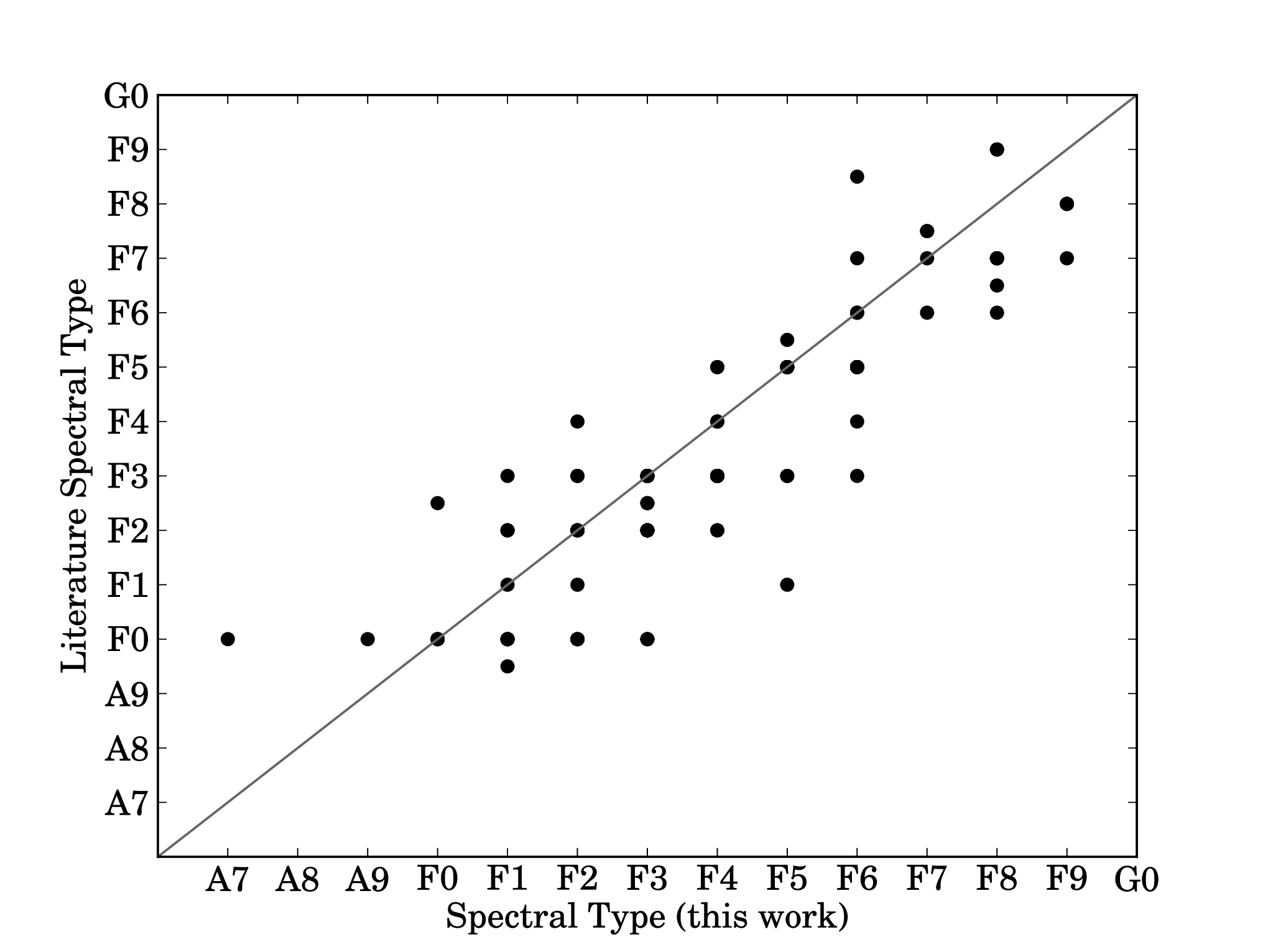}
\caption{
\label{fig:spt_compare}
  Comparison of our newly obtained spectral types with those available in the literature:
  \cite{1988mcts.book.....H, 1982mcts.book.....H,1975mcts.book.....H, 1978mcts.book.....H}.  
  A line with unit slope is plotted to guide the eye.
}
\end{center}
\end{figure}

Spectral types for many of our candidate members are available through the 
Michigan Spectral Survey 
\citep{1988mcts.book.....H,1982mcts.book.....H,1978mcts.book.....H,1975mcts.book.....H}
which allows us to directly compare our newly obtained spectral types with past results 
from plate surveys.  As Figure~\ref{fig:spt_compare} shows, the newly obtained spectral
types agree quite well with those obtained in the Michigan Spectral Survey.  The average
difference is small: 0.5 $\pm$ 1.1 (1$\sigma$).  
The Houk classifications are based on comparison to the ``MK'' standards
from \citet{1953ApJ...117..313J}, whereas we have relied heavily on the
``revised MK'' F-type standards from \citet[``MK78'']{1978rmsa.book.....M} 
(these later types have been adopted by Gray, Garrison, Corbally and 
collaborators in their surveys). Any systematic
differences in the classifications likely come from differences in the
choice of standard star, which differ between us and Houk mostly among
the earliest and latest F stars.

\subsection{Extinction\label{sct:extinction}}
To calculate individual extinctions in the Johnson-Cousins and 2MASS
bands, we constructed a modern spectral type-intrinsic color
calibration with B-V, V-I$_{c}$, V-K$_S$, J-H, and H-K$_S$ colors
using samples of nearby dwarf stars from the Hipparcos catalog
\citep[][]{1997ESASP1200.....P}, normal cool dwarfs from Neill Reid's
compiled photometric catalog for stars in the 3rd Catalog of Nearby
Stars\footnote{http://www.stsci.edu/$\sim$inr/cmd.html}, and a
cross-matched catalog combining the new Gliese-2MASS catalog
\citep{2010PASP..122..885S} and the catalog of average Johnson UBV photometry
from \citet{2006yCat.2168....0M}.  For the Hipparcos sample, only dwarfs with
luminosity class ``V'' and M$_V$ within 1 mag of the \citet{2005AJ....129.1776W}
main sequence were selected, and only stars with
parallaxes greater than 13.33 mas and relative parallax error less
than 12.5\% were included (d $<$ 75 pc), to select those stars
ostensibly within the ``Local Bubble'' which has negligible reddening.
An intrinsic color sequence was constructed for B through M-type
stars, and the adopted intrinsic colors for a given dwarf spectral
subtype were adopted based on a best B-V color\footnote{\label{ftn:spt}Detailed notes 
for each subtype are available in the files listed at 
http://www.pas.rochester.edu/$\sim$emamajek/spt/ 
or http://www.ctio.noao.edu/$\sim$emamajek/spt/}. 
Our adopted intrinsic colors are in Table~\ref{tbl:ic_teff}.

Many of the stars in our sample had very low extinction, and where we obtained
a non-physical negative extinction we set the extinction to zero.  We used a
total-to-selective extinction of R$_V$=3.1 and calculated A$_V$ using the color excesses E(B-V), 
E(V-I$_c$), E(V-J), E(V-H) and E(V-K$_S$) with the extinction ratios A$_{I_c}$/A$_V$=0.58, 
A$_J$/A$_V$=0.27, A$_H$/A$_V$=0.17, A$_{K_S}$/A$_V$=0.11 \citep{2003A&A...401..781F} 
to estimate extinctions for each star.  We adopted the median A$_V$ with the 
standard deviation as a conservative estimate of the uncertainty.  These are 
listed in Table~\ref{tbl:stellarparams} with our other derived stellar
parameters.

\subsection{Distances \label{sct:dist}}
While the stars in our sample have been selected from the {\it Hipparcos} catalog and
therefore have trigonometric parallaxes, we can refine the distance determination
by employing moving cluster, or ``convergent point'' parallaxes.  This works by leveraging 
the longer-baseline astrometric measurements employed to determine proper motions and 
taking advantage of the common space motion for the group.  Kinematically improved parallaxes 
have been shown to reduce scatter in the H-R diagram (see the Hyades in 
\cite{2001A&A...367..111D} for a good example) and a comparison with trigonometric 
parallaxes can further help identify interlopers.  One caveat is that 
kinematic parallaxes are meaningless for stars which are not true members.

\begin{figure}
\begin{center}
\includegraphics[scale=0.45]{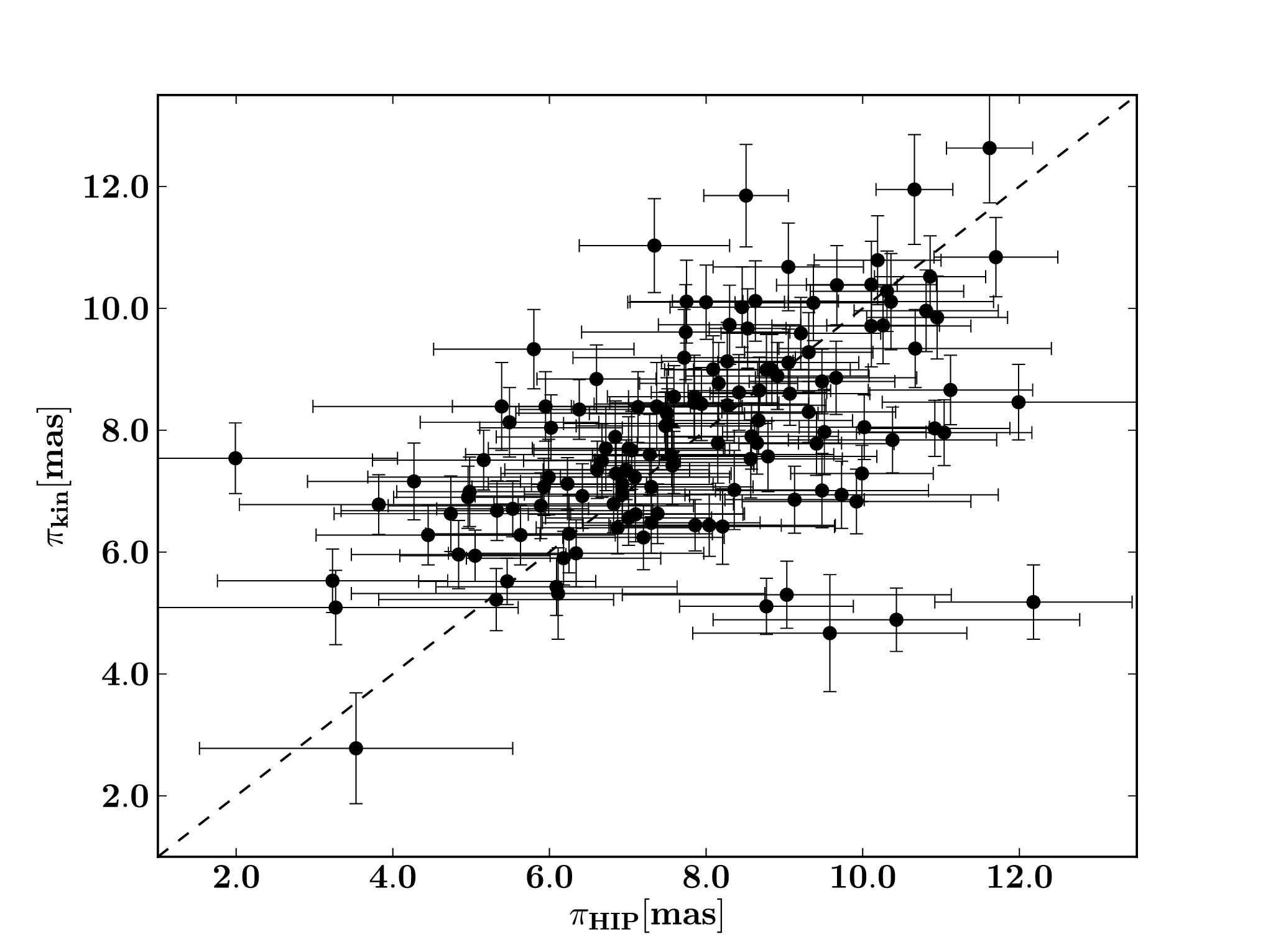}
\caption{
\label{fig:plx}
  Comparison between {\it Hipparcos} trigonometric parallaxes and our kinematic parallaxes.
  A line with unit slope is plotted for comparison.
}
\end{center}
\end{figure}

For our kinematic parallax calculations we followed \citet{2005ApJ...634.1385M} and used
updated space motions for US, UCL and LCC tabulated in \cite{2011ApJ...738..122C}\footnote{
These updated space motions should yield the best available predicted radial velocities 
and kinematic parallaxes.}.  As described 
previously, we use proper motion data from the revised {\it Hipparcos} 
reduction \citep{2007A&A...474..653V} for stars with single star (five-parameter) solutions 
and data from Tycho-2 \citep{2000A&A...355L..27H} otherwise.  Our kinematic parallaxes 
compare very well with the {\it Hipparcos} trigonometric parallax data (Figure~\ref{fig:plx}).  

\subsection{Field Star Contamination}
While the techniques used to define the sample in \citet{1999AJ....117..354D} ensure
that the candidate members are physically coincident with Sco-Cen and moving with similar
space motion, the selection is purely based on astrometric information.  Here we attempt
to use our spectroscopic observations in conjunction with an additional kinematic criterion
to identify interlopers.

While the strength of Li can be used as a youth indicator in G- and K-type stars,
older F-type stars will show less Li depletion and thus we cannot rely on this spectral 
feature alone as a youth indicator in our sample (\citealp[][however see]{1991MmSAI..62...33B}
\citealp{2011ApJ...738..122C}).  We can, however, use Li
to determine membership for any G/K stars which are companions to F-type stars, excluding or
confirming the {\it pair} as members.  Since our sample should consist of stars with 
predominantly dwarf and subgiant like gravities, we can also identify and exclude any giants found in our sample.

For members, the kinematic parallaxes should be in agreement with the
trigonometric parallaxes.  As a conservative criterion, we reject any star as a member if the
difference between trigonometric parallax and kinematic parallax exceeds three 
times the uncertainty in those quantities added in quadrature.  The motivation 
is that if the disagreement between the trigonometric and kinematic parallaxes is significant at the 3$\sigma$ level,
the object clearly has a different space motion than the group and can be safely rejected.

\subsubsection{US Interlopers}
Using our conservative kinematic rejection criteria, we rule out HIP~79258 as a member.  A 
kinematic parallax of 5.11$\pm$0.46~mas disagreed with the trigonometric parallax of 8.77$\pm$1.11~mas at 
a confidence level $>3\sigma$.  This is the only interloper we were able to identify in Upper Sco.
We then expect that the US sample of 21 still contains $\sim$4 interlopers, $\sim$19\%.

\subsubsection{UCL Interlopers}
\citet{2011ApJ...738..122C} reject HIP~70833 as a member of UCL since the cooler K3IV companion has an 
EW(Li~$\lambda$6707$\AA$) of only 0.01~\text{\AA}.  \citet{2011ApJ...738..122C} also reject HIP~75824 
based on its low Li, with EW(Li~$\lambda$6707$\AA$)$<$8~\text{\AA}.  
Finally, HIP~69327 lies far below the zero-age main sequence, an un-physical portion of the H-R diagram, 
when the kinematic distance is used.  
As this is inconsistent with group membership, we identify it as an interloper.

With these non-members identified in UCL, there still remain $\sim$12 interlopers in our 
UCL sample of 53, or $\sim$23\%.

\subsubsection{LCC Interlopers}
We reject one giant as a LCC member (HIP~62428, A7III; also rejected by \citealt{2011ApJ...738..122C}). 
It was also classified as a giant by \citet{1975mcts.book.....H}.
\citet{2011ApJ...738..122C} also exclude HIP~66285 as a member, based on the lack of Li in its
cooler co-moving companion.
HIP~59781 has a kinematic parallax of 5.18$\pm$0.61~mas and a trigonometric parallax
of 12.58$\pm$1.26~mas, disagreeing above our 3$\sigma$ threshold so we reject it.
HIP~56227 has a kinematic parallax of 11.85$\pm$0.84~mas and a trigonometric parallax
of 8.51$\pm$0.54~mas, also disagreeing above our 3$\sigma$ threshold so we reject it.
HIP~57595 has a trigonometric parallax of 3.53$\pm$2.00 mas, which places it 165~pc from the
mean distance of LCC.  The Tycho-2 proper motion for this star supports
it being an unrelated background star -- the kinematic parallax assuming it were a member
of LCC places even further from LCC ($\pi_{kin}$=2.78$\pm$0.91 mas), so we reject it.
Finally, HIP~63022, HIP~61086, HIP~64316, HIP~62674, and HIP~62056 occupy an un-physical 
portion of the H-R diagram (below the zero-age main sequence) when the kinematic
distances are used.  As this is inconsistent with group membership, we identify 
these objects as interlopers.

With these interlopers identified in LCC, we believe that few, if any, of the remaining
F-type stars are interlopers.  Stars rejected as Sco-Cen members are listed in 
Table~\ref{tbl:rejected}.

\subsection{Accretion Disk Fraction}
We can examine our red optical spectra for evidence of accretion
with the H{\rm $\alpha$} feature.  Accreting stars are thought to produce H{\rm $\alpha$} 
emission due to hot infalling gas.  For our accretion criterion we use the 
H{\rm $\alpha$} full width at 10\% peak (W$_{10}$(H$\alpha$)).  \cite{2003ApJ...582.1109W} find 
that independent of spectral type, a full width at 10\% peak $>$ 270 km~s$^{-1}$ is a good 
indicator of accretion.  We do not estimate the mass accretion rates as that is is beyond 
the scope of this study.

\begin{figure}
\begin{center}
\includegraphics[scale=0.45]{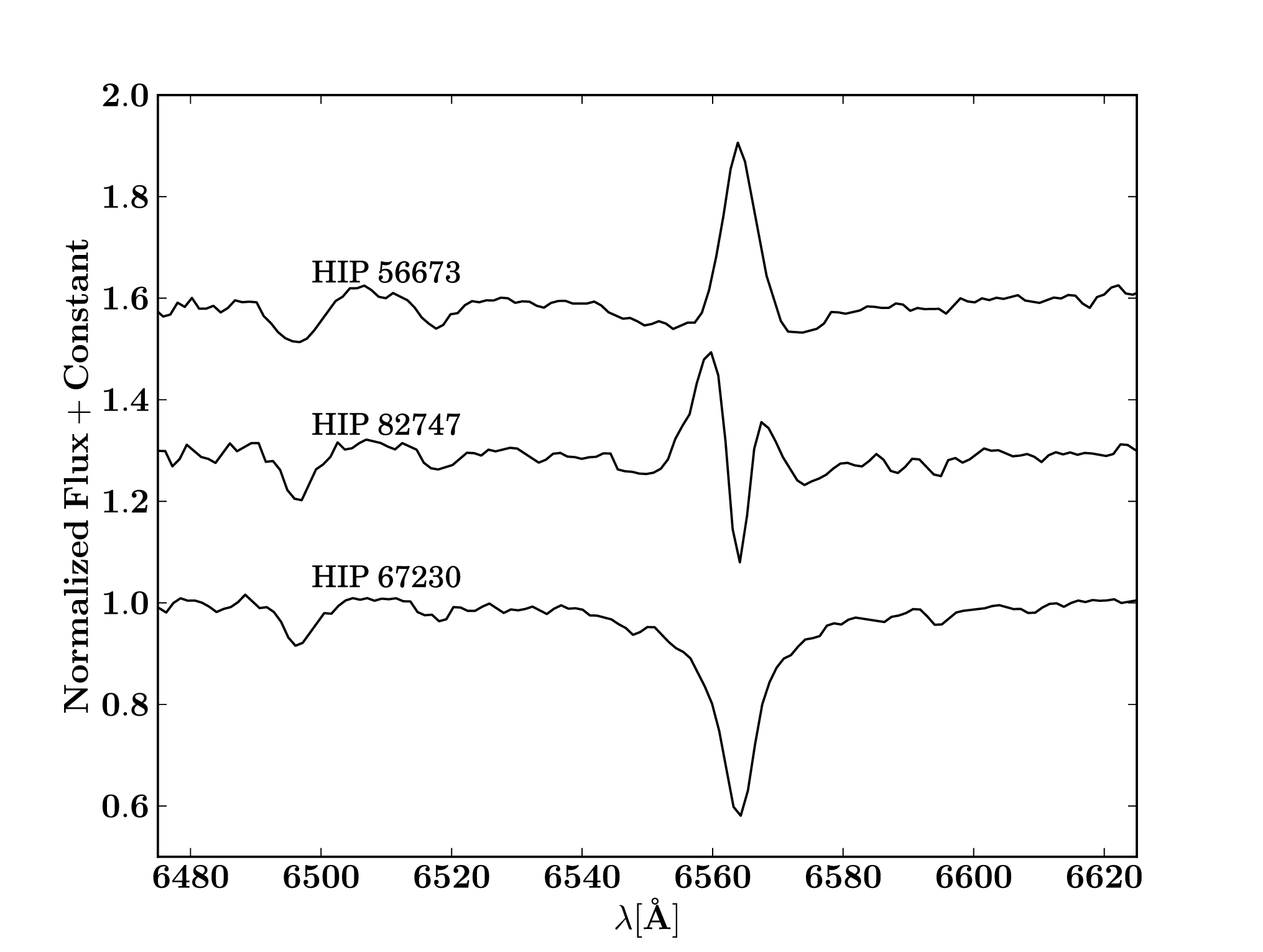}
\caption{\label{fig:halpha}
 H$\alpha$ region showing the two accretors in our sample, HD~101088 (HIP~56673, F5IVe) and AK~Sco (HIP~82747, F5Ve),
 along with a non-accreting star (HIP~67230, F5V) of the same spectral type.
}
\end{center}
\end{figure}

Using this criterion, we find only two stars in our sample with H$\alpha$ emission consistent 
with accretion, both of which are binaries: the near-equal mass system AK Sco
\citep[HIP~82747;][]{1989A&A...219..142A,2003A&A...409.1037A}
and the recently studied HD~101088 \citep[HIP~56673;][]{2010ApJ...714.1542B} with 10\% peaks
widths of 710~km~s$^{-1}$ and 400~km~s$^{-1}$, respectively.  The spectral resolution
of our red optical spectra at H$\alpha$ is $\sim$140~km~s$^{-1}$.  The H$\alpha$ profiles of 
these accretors are shown in Figure~\ref{fig:halpha} along with the H$\alpha$ profile of a 
non-accreting star in our sample (HIP~67230, F5V) with the same spectral type as HD~101088 
and AK~Sco.

\begin{figure}
\begin{center}
\includegraphics[scale=0.45]{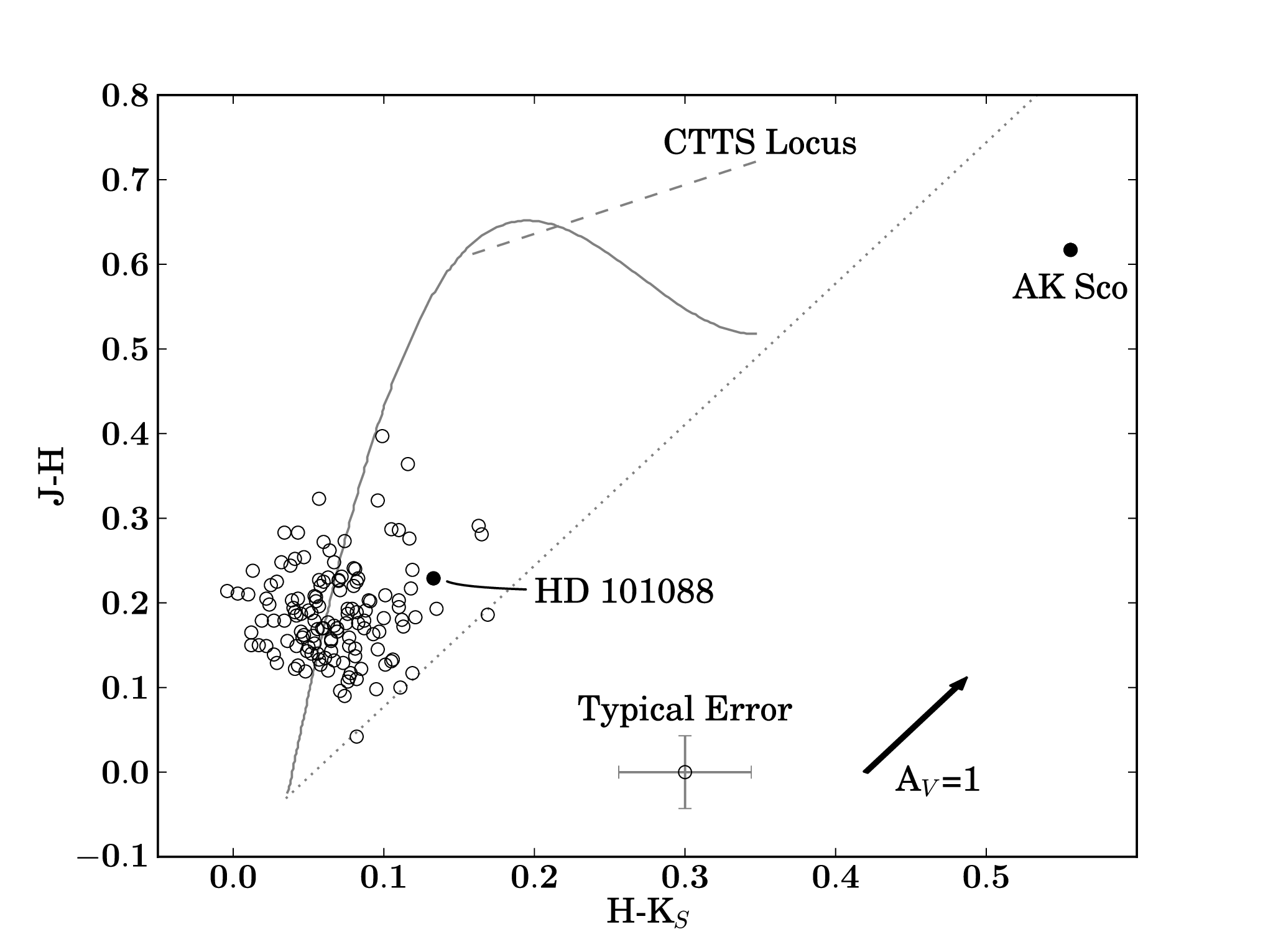}
\caption{\label{fig:color-color}
 H-K$_S$ vs. J-H for the program stars (not de-reddened; excluding interlopers).  The solid line
 is the dwarf color locus and the dashed line is the classical T Tauri locus of 
 \cite{1997AJ....114..288M}, shown for reference.  The dotted line is the reddening line of a
 standard A0 star.  The solid circle outlier to the right is the eclipsing binary AK~Sco, indicating
 that AK~Sco's NIR color excess cannot be due to reddening.
}
\end{center}
\end{figure}

Using the classical T Tauri star color-color locus as determined by \citet{1997AJ....114..288M}
and the intrinsic color locus of main-sequence stars we look for near-IR photometric signatures of accretion
among our Sco-Cen members using {\it 2MASS} near-IR photometry.  The plot in Figure~\ref{fig:color-color} shows our Sco-Cen members 
(without reddening corrections applied) clustered around the locus with spread consistent with the 
uncertainties in the colors.  The one obvious outlier with H-K$_S$ $>$ 0.4 is 
the known near-equal mass binary AK~Sco referred to above.  AK~Sco is the star to the right of 
the locus, and its position is similar to that of the Herbig Ae/Be stars in \cite{2005AJ....129..856H}.

Here we estimate the optical spectroscopic accretion disk fraction for F stars 
in our sample as 0/17 ($<$19\%; 95\% C.L.) for US, while UCL has 1/41 
($2^{+5}_{-1}$\%; 68\% C.L.) accretors and LCC has 1/50 ($2^{+4}_{-1}$\%; 68\% C.L.) 
F-type accretors.  This compares well with the \cite{2006ApJ...651L..49C}
{\it Spitzer} results in which 0/30 ($<$11\%; 95\% C.L.) F- and G-type stars 
were found to be accretors. The F-type results in UCL and LCC are consistent 
with the lower-mass G- and K-type stars studied by \citet{2002AJ....124.1670M}, 
in which 1/110 ($0.9^{+2.0}_{-0.3}$\%; 68\% C.L.) members exhibited spectroscopic 
evidence of accretion.

\begin{figure}
\begin{center}
\includegraphics[scale=0.45]{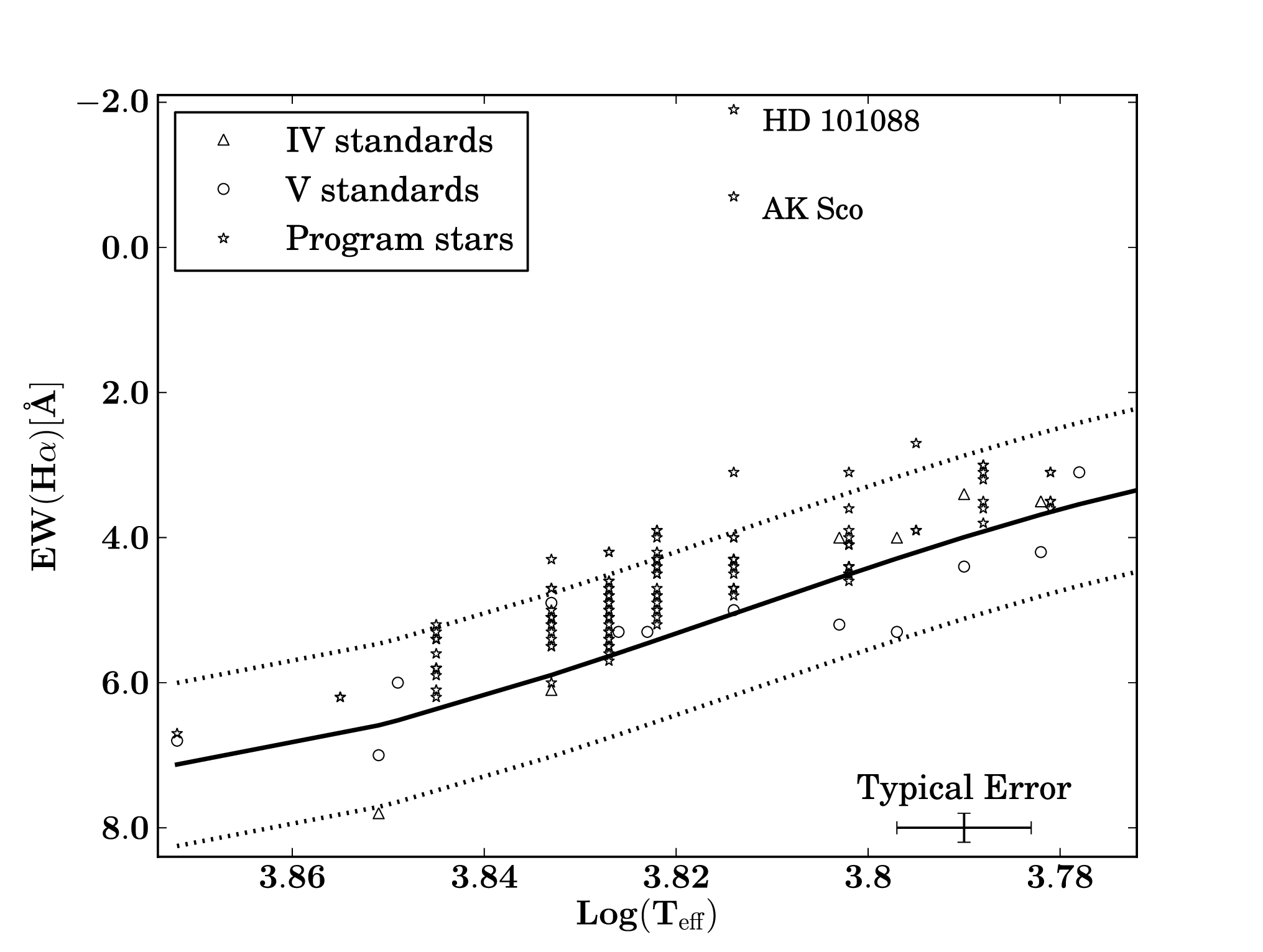}
\caption{\label{fig:activity_teff}
  H$\alpha$ EWs for the member stars compared with inactive field dwarfs and subgiants.
  The solid line is a third order polynomial fit to the field stars and the dotted lines 
  represent the 2$\sigma$ scatter in the fit to the field stars.  While the Sco-Cen
  program stars (interlopers not shown) are heavily represented above the trend line, 
  most of them still lie within of the 2$\sigma$ spread of the field stars.
}
\end{center}
\end{figure}

As mentioned previously, pre-main sequence stars may exhibit enhanced chromospheric 
activity, which will manifest itself through partially filled H$\alpha$
absorption lines.  To examine this effect we measured the strength of the H$\alpha$ lines
with IRAF\footnote{IRAF is distributed by the National Optical Astronomy Observatory, 
which is operated by the Association of Universities for Research in Astronomy (AURA) 
under cooperative agreement with the National Science Foundation.} 
for all our program stars as well as our spectral standards and 
plot the EW(H$\alpha$) against \teff\, (Figure~\ref{fig:activity_teff}).
The members plotted here indicate slightly enhanced 
chromospheric activity, although most are still within the 2$\sigma$ spread of the 
spectral standards. 

\subsection{H-R Diagram\label{sct:hrd}}
In order to compare Sco-Cen members with theoretical models and explore the star formation
history of the F-type members, we construct a theoretical H-R diagram.  The adopted effective
temperature (\teff) scale and bolometric correction (BC) scale was taken from an
extensive set of notes for spectral subtypes by EM\footnotemark[6].  This new 
\teff\,  scale comes from careful review and inter-comparison of
spectroscopic and photometric temperature estimates for high quality
MK standards as well as large samples of field dwarfs with MK classifications. The bolometric corrections 
are consensus estimates for the adopted temperatures, which rely heavily on the BCs for hot dwarf stars from
\citet{1998A&A...333..231B} and for cool dwarf stars on the series of papers by
\citet{2006MNRAS.373...13C, 2008MNRAS.389..585C, 2010A&A...512A..54C} (however BCs were
also calculated interpolating the scales of \citet{1976ApJ...203..417C}, 
\citet{1994MNRAS.268..119B}, \citet{1996ApJ...469..355F}, and 
\citet{2004AJ....128..829B} for comparison).  Our temperatures and bolometric 
corrections are listed in Table~\ref{tbl:ic_teff}.

\begin{figure}
\begin{center}
\includegraphics[scale=0.45]{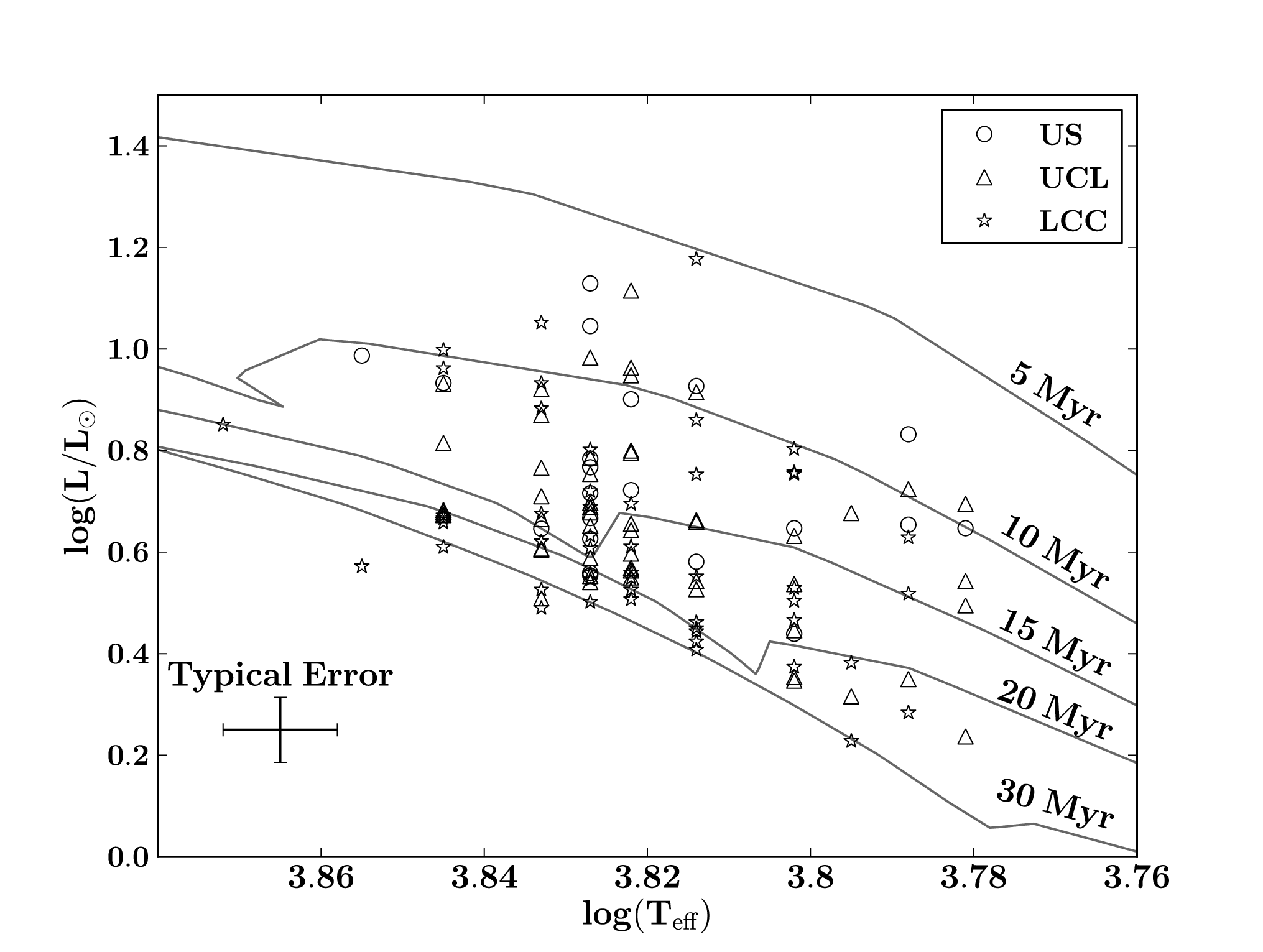}
\caption{\label{fig:hrd_fstars}
  H-R diagram for all stars except identified interlopers.  Circles,
  triangles and star symbols are US, UCL and LCC candidate members,
  respectively.  Plotted for comparison are 5, 10, 15, 20 and 30 Myr isochrones
  from the \citet{2008ApJS..178...89D} models.
}
\end{center}
\end{figure}

We combine our extinction estimates with our kinematic parallaxes and 
V-band photometry \citep{1997ESASP1200.....P} to estimate bolometric 
luminosity and place the members on a theoretical H-R diagram.  We compare our data to the 
Dartmouth PMS models \citep{2008ApJS..178...89D} in Figure~\ref{fig:hrd_fstars}.

\section{Discussion}
\begin{figure}
\begin{center}
\includegraphics[scale=0.45]{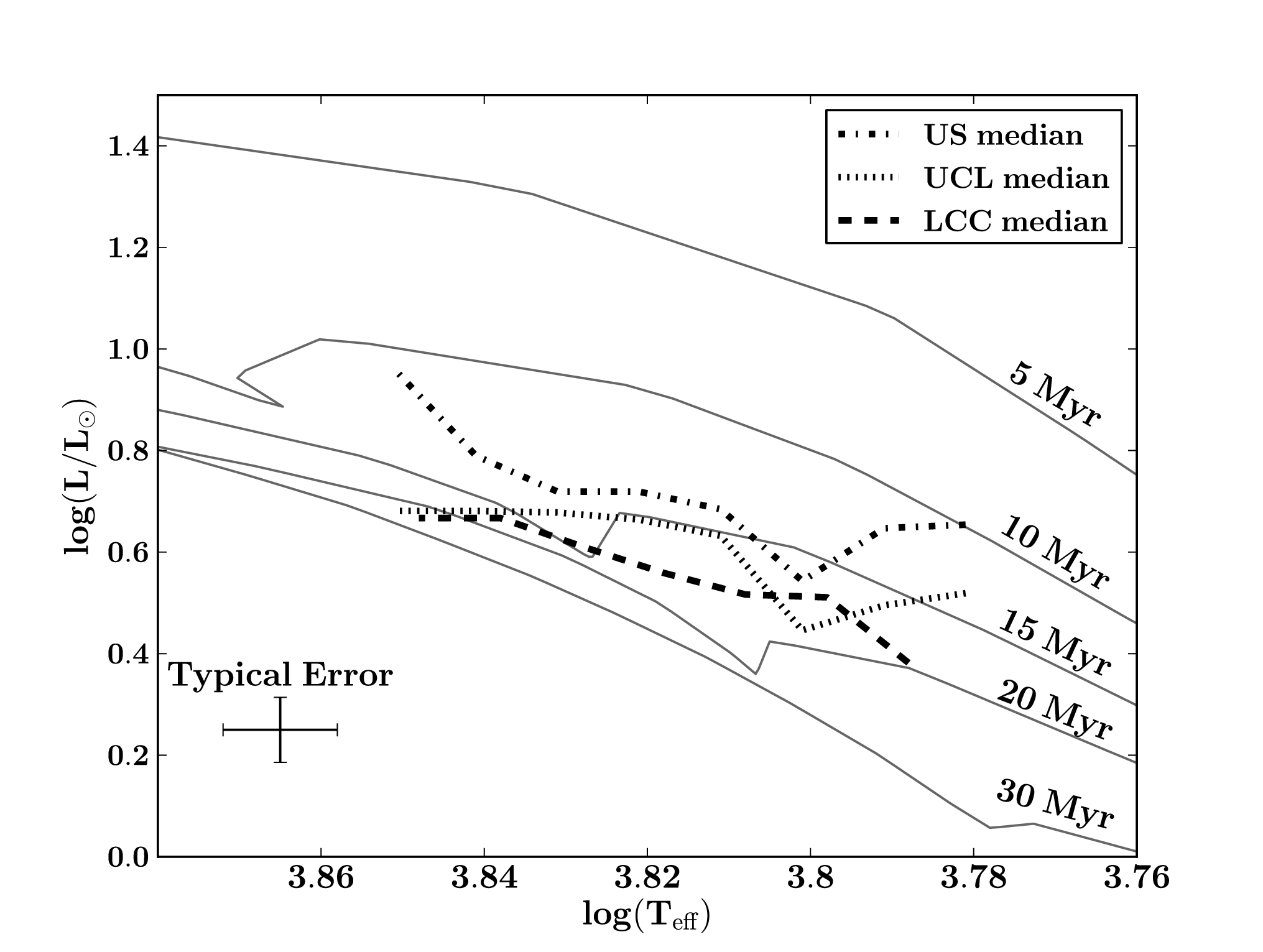}
\caption{\label{fig:hrd_fstars_iso}
  H-R diagram with empirical isochrones for US (dot-dashed),
  UCL (dotted) and LCC (solid), constructed by taking the median
  luminosity over a 0.01 dex \teff\, step size with a 
  0.025 dex log(\teff) window.  Plotted for comparison are
  the \citet{2008ApJS..178...89D} models.
}
\end{center}
\end{figure}

Immediately noticeable in the H-R diagram (Figure~\ref{fig:hrd_fstars})
is that the Upper Sco F-stars are not clustered around the 5~Myr isochrone, the age
often quoted for US \citep{1989A&A...216...44D,2002AJ....124..404P}.  In fact if a
``moving median'' empirical isochrone  is plotted (Figure~\ref{fig:hrd_fstars_iso}), it is $\sim$0.4 dex 
below the 5~Myr theoretical isochrone from the \citet{2008ApJS..178...89D} models.  
While a choice of different theoretical tracks vary this disagreement slightly 
(e.g., \citet{1997MmSAI..68..807D}, \citet{2000A&A...358..593S}, 
\citet{2004ApJS..155..667D}) the disparity is at least $\sim$0.4 dex at spectral type F3, which 
corresponds to a factor of $\gtrsim$ 2.5 in luminosity, or about 1 mag.  This is such a large 
effect that it clearly cannot be attributed to the uncertainty 
in the photometry or the errors in spectral types.  

Although there is some scatter, the positions of the empirical isochrones in 
Figure~\ref{fig:hrd_fstars_iso} are consistent with the age rank from previous results 
\citep{2002AJ....124.1670M}; from oldest to youngest: LCC, UCL, US.  A close examination
of the empirical isochrones reveals that UCL appears to be reaching the main sequence
at log(\teff) $\simeq$3.84 or spectral type $\sim$F2, while LCC appears to be reaching the 
main sequence at log(\teff) $\simeq$3.82 or spectral type $\sim$F4.  
The F-type members of US appear to be all pre-main sequence, and thus we do not
see a main sequence turn-on point for US among F-type stars.

\subsection{Ages}
\begin{figure}
\begin{center}
\includegraphics[scale=0.45]{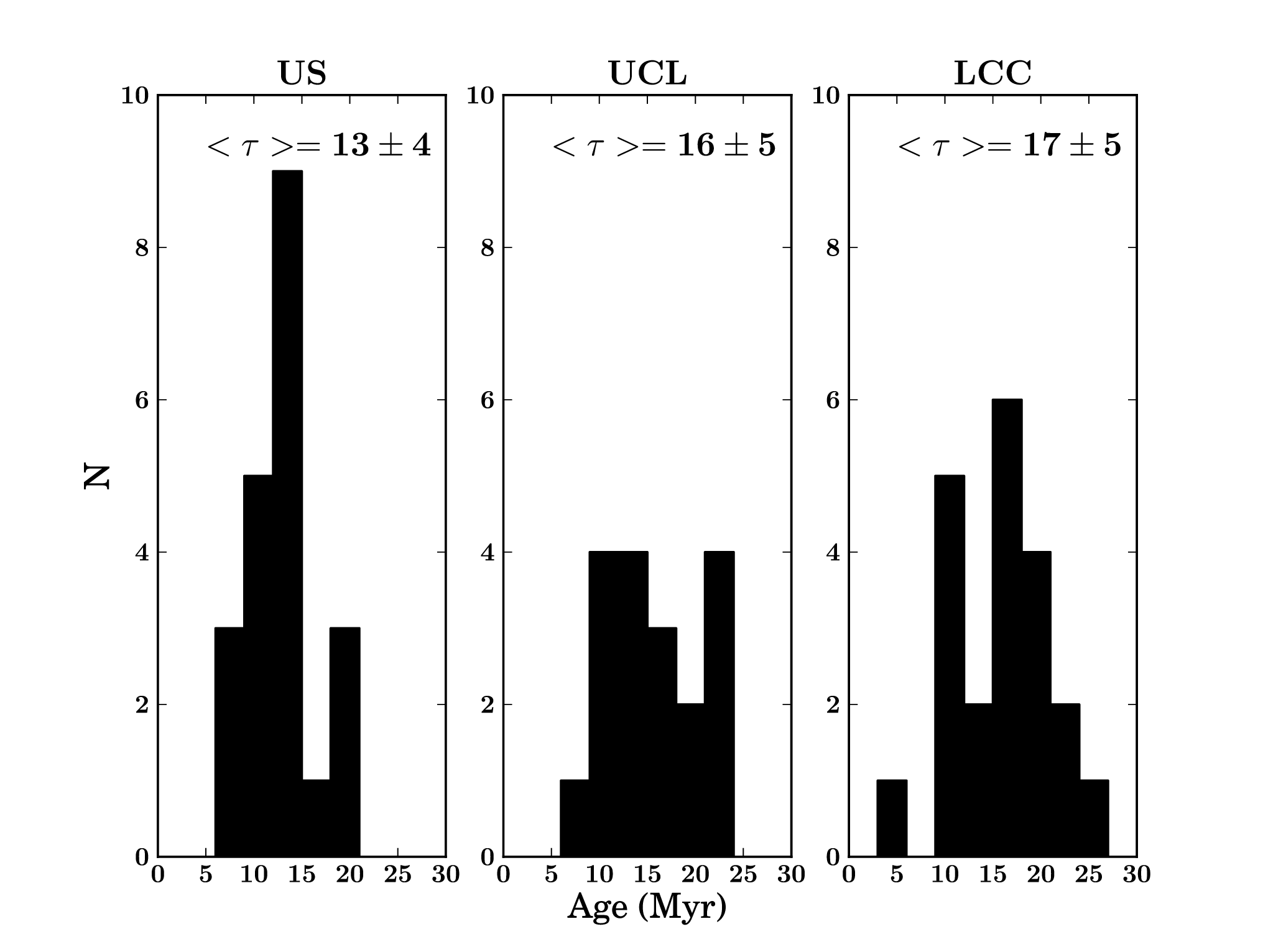}
\caption{\label{fig:age_hist}
  Distribution of ages obtained with the \citet{2008ApJS..178...89D} evolutionary
  models.  For UCL and LCC, only stars cooler than F5 were considered.
}
\end{center}
\end{figure}

For stars in UCL and LCC earlier than $\sim$F5, the H-R diagram positions of the stars 
are very near the main sequence and thus we can not reliably determine ages from
pre-main sequence evolutionary tracks.  Therefore, we only use F5 and cooler stars to estimate
the median age of UCL and LCC.  For all stars in US and those later than F5 in UCL and LCC 
we calculate ages and masses by linearly interpolating between isochrones from 
\citet{2008ApJS..178...89D}, \citet{2004ApJS..155..667D}, \citet{2000A&A...358..593S}, 
and \citet{1997MmSAI..68..807D}.  Figure~\ref{fig:age_hist} shows the distribution of 
ages for the three subgroups.  Individual results for our ages and masses are listed in 
Table~\ref{tbl:stellarparams}.  For stars which are too close to the main sequence to 
infer a reliable age, we determine a lower limit on the age using the 2$\sigma$ upper 
limit on the luminosity and estimate a mass by assuming an age of 15~Myr (10~Myr for US).
Our median age estimates for each subgroup are listed in Table~\ref{tbl:ages}.

Interestingly, we obtain a significantly older median age for Upper Sco than other studies
of the low-mass members (5~Myr, \citealt{2002AJ....124..404P}; 6-8~Myr, \citealt{1985ASSL..120...95D}; 
10~Myr, \citealt{1971PhDT.........3G}).  As seen in Figure~\ref{fig:age_hist},
we obtain a median age of 13~Myr for Upper Sco using the evolutionary tracks of 
\citet{2008ApJS..178...89D} and an overall median age from all tracks of 13$\pm$1~Myr (68\% C.L.).
This is disconcerting, especially considering the median F-star 
ages we obtain for LCC and UCL are consistent with the main sequence turn-off ages and the 
pre-main sequence G- and K-type stars examined in \citet{2002AJ....124.1670M}.  Motivated 
to resolve this large discrepancy, we revisit the Upper Sco main sequence turn-off ages,
the age of the M supergiant Antares, examine the ages of the A-type stars and the ages 
for the pre-main sequence G-type stars.
We also examine a kinematic expansion age for Upper Sco.

\subsubsection{Intrinsic Age Spreads}
In order to constrain how much of the observed spread in ages is due to observational
uncertainty and estimate the intrinsic age spread, we performed a simple Monte Carlo
simulation.  First we calculated the observed 1$\sigma$ dispersion in ages for each subgroup
by simply taking the standard deviation of the ages (F5 and later stars for UCL and LCC),
with deviant outliers clipped using Chauvenet's criterion \citep{2003drea.book.....B}.
These observed spreads are shown in Table~\ref{tbl:ages}.  

\begin{figure}
\begin{center}
\includegraphics[scale=0.45]{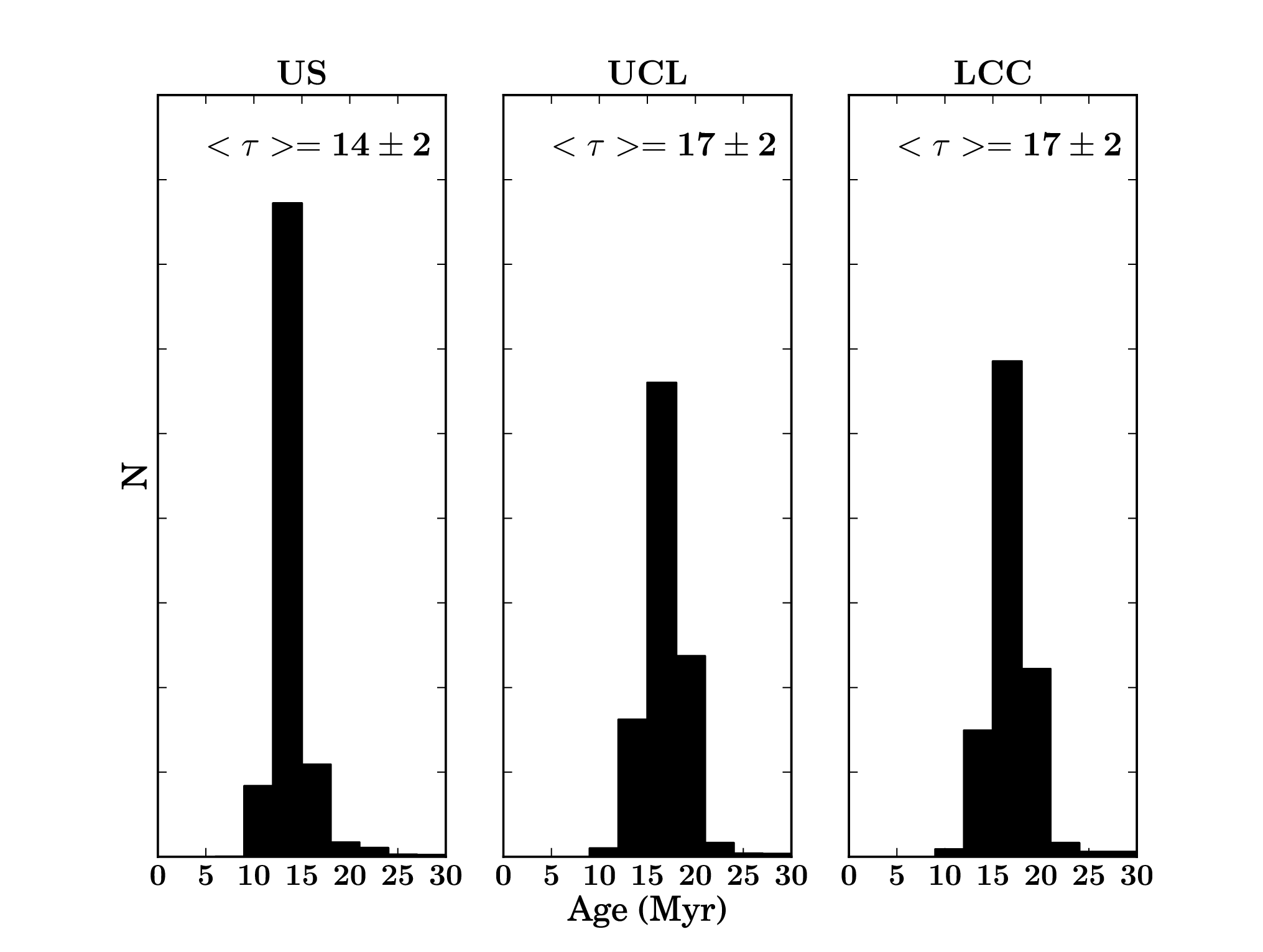}
\caption{\label{fig:simages}
  Distribution of ages obtained from a simulated population of 10$^4$ stars with 
  median H-R diagram position and uncertainty with a Gaussian distribution (for 
  UCL and LCC only members F5 or later were used).  Ages (in Myr) were obtained using the
  \citet{2008ApJS..178...89D} models.
}
\end{center}
\end{figure}

For each subgroup, we generated a synthetic population of 10$^4$ H-R diagram positions using the 
median H-R diagram position and median uncertainties for our member stars (F5 or later only for 
UCL and LCC), assuming a Gaussian distribution of errors.  We then obtained ages and calculated 
the 1$\sigma$ spread in ages for these three synthetic populations,
again clipping deviant outliers with Chauvenet's criterion since the same criterion was applied to 
our real data to obtain the 1$\sigma$ dispersion in ages. The distribution of ages for these 
simulated populations with the \citet{2008ApJS..178...89D} evolutionary tracks are shown in 
Figure~\ref{fig:simages}.

Our estimates here assume that the observed dispersion in ages is the result
of the true dispersion in ages and the dispersion in ages as a result of the
observational uncertainties, i.e. $\sigma_{observed}^2 = \sigma_{intrinsic}^2 + 
\sigma_{uncertainties}^2$.  Using the \citet{2008ApJS..178...89D} models, we
then estimate the intrinsic dispersion in ages using the dispersions from our
real and simulated populations.  Rather than give preference to one set of evolutionary
models, we repeat this calculation for each of our evolutionary models.  This 
way we obtain a range of intrinsic age dispersions.  

For US the observed 1$\sigma$ dispersion in ages ranges from $\pm$3~Myr with the 
\citet{1997MmSAI..68..807D} models to $\pm$5~Myr with the \citet{2000A&A...358..593S} 
models.  When we account for the age dispersion from the observational uncertainties, 
we find the intrinsic 1$\sigma$ age dispersion ranges from $\pm$1~Myr to $\pm$3~Myr, again 
depending on the models.  For UCL and LCC, which have an observed 
age dispersion of $\pm$5--9~Myr and $\pm$5--9~Myr (depending on models), we find 
intrinsic age dispersions of $\pm$4--7~Myr and $\pm$0--8~Myr, respectively.  This 
indicates that 68\% of the star formation in Sco-Cen occurred over a 
time span of 2--6~Myr (US), 8--14~Myr (UCL) and $<$16~Myr (LCC).  These are upper 
limits as we have not accounted for stellar binarity.  Our findings 
show a much smaller age dispersion for US than the other two subgroups, 
consistent with the smaller age spreads found by \citet{2002AJ....124..404P} 
and \citet{2008ApJ...688..377S}, who also found a very small intrinsic dispersion 
in age.  For UCL and LCC our age spreads are larger than that found in  
\cite{2002AJ....124.1670M}, who found that 68\% of the star formation had occurred within
4--6~Myr for UCL and LCC. 

\subsection{Upper Sco Main Sequence Turn-Off Revisited}
The most recent age determination for the turn-off was performed by 
\citet{1989A&A...216...44D} using Walraven photometry in which they 
estimated a turn-off age of 5~Myr using \cite{1981A&A...102..401M}
evolutionary tracks with overshooting but no mass loss or rotation.  \citet{2002AJ....124..404P} 
examined the H-R diagram again and argued that the data were consistent 
with an age of 5~Myr with very little spread in ages.  We now have 
updated stellar evolutionary models and {\it Hipparcos} astrometry 
which we employ to revisit the turn-off age.  We use the early-type members from
\citet{1999AJ....117..354D} with the addition of $\delta$~Sco, a long-period binary that is 
certainly a member of US but was not a \cite{1999AJ....117..354D} member of US due to its 
perturbed motion.  To construct the H-R 
diagram, we take Str\"{o}mgren photometry from \citet{1998A&AS..129..431H} and de-reddened
it according to the prescription of \citet{1983MNRAS.205.1215S}.  We then use the 
\teff\, and BC$_V$ calibration of \citet{1994MNRAS.268..119B} with
the de-reddened Str\"{o}mgren photometry.  Since we are using the spread in isochrones
near where the stars evolve off the main sequence, we restrict ourselves to stars hotter 
than log(\teff) $>$ 4.40 dex or more luminous than log(L/\lsun) $>$ 4.00 dex.  For stars cooler or less 
luminous than this the derived ages are very sensitive to the observational uncertainties and we 
are unable to estimate meaningful ages.
Our derived temperatures and luminosities are given in 
Table \ref{tbl:usturnoff}.  A plot of the data, with and without corrections for binarity (see below), 
are shown in Figure~\ref{fig:usturnoff} with the evolutionary tracks of \citet{1994A&AS..106..275B}.

We have not used the runaway star $\zeta$ Oph in our analysis, for several reasons.  First,
while \citet{2000ApJ...544L.133H} used {\rm Hipparcos} astrometry to conclude that $\zeta$~Oph
most likely originated from Upper Sco $\sim$1~Myr ago, they note that it could have also 
originated from UCL $\sim$3~Myr ago.  Unfortunately, the radial velocity of this star is 
poorly constrained\footnote{+15 km~s$^{-1}$, \cite{1993ApJ...417..320R}; 
+6 km~s$^{-1}$, \cite{1980ApJ...242.1063G}; -12.6 km~s$^{-1}$, 
\cite{1977ApJ...214..759C}; -15 km~s$^{-1}$, \cite{2004ApJS..152..251V}} and thus the UVW space 
motion is not sufficiently constrained to associate it with Upper Sco with great confidence.  More
importantly, however, close examination of the abundance patterns in $\zeta$~Oph 
indicate that it has an anomalously high helium abundance \citep{1992A&A...261..209H} and thus 
likely participated in a mass transfer event, perhaps from a close former binary companion.  
For this reason it is not reliable to use it to determine the turn-off age since it may have
received extra nuclear fuel unaccounted for by the evolutionary models.

With only six B-type stars determining the turn-off age, it is important to assess the 
binarity of these stars since it may significantly alter the H-R diagram positions and, 
therefore, the derived ages. 
Only two stars, $\omega$~Sco and $\tau$~Sco, do not have a detected companion.  
$\beta^1$~Sco is a binary with $\Delta V$=1.26$\pm$0.17 \citep{1976ApJ...207..994E}.  
This magnitude shift, when accounted for, moves the primary down on the H-R diagram $\sim$0.1 dex
in luminosity and gives an age for the primary of 9~Myr.
$\pi$~Sco is a binary with $\Delta V \sim$3.7 \citep{1996Obs...116..387S}, 
which moves the primary down on the H-R diagram $\sim$0.01 dex
in luminosity, having virtually no effect on the derived age for $\pi$~Sco.
$\delta$~Sco is a binary with $\Delta V$=1.5-1.9 \citep{2009MNRAS.396..842T}, 
shifting the primary down $\sim$0.07-0.1 dex in luminosity.  However, because of 
its position it effectively moves parallel to the 9~Myr isochrone and
thus does not affect the derived age for the primary.
$\sigma$~Sco is a quadruple system discussed extensively by \citet{2007MNRAS.380.1276N}.  They report 
the system as a spectroscopic pair (which includes the primary), a tertiary B7 component and a B9.5V common 
proper motion companion separated by 20$"$.  They deconstructed the spectroscopic pair into a B1III primary 
with a B1V secondary with $\Delta V \simeq$ 0.80, and a system age of $\sim$10~Myr using the \citet{2004A&A...424..919C} 
evolutionary models.  Correcting our derived H-R diagram position for the magnitude of the secondary, we
find that the primary moves down $\sim$0.2 dex, which changes our derived age to 8~Myr for the primary.

\begin{figure}
\begin{center}
\includegraphics[scale=0.45]{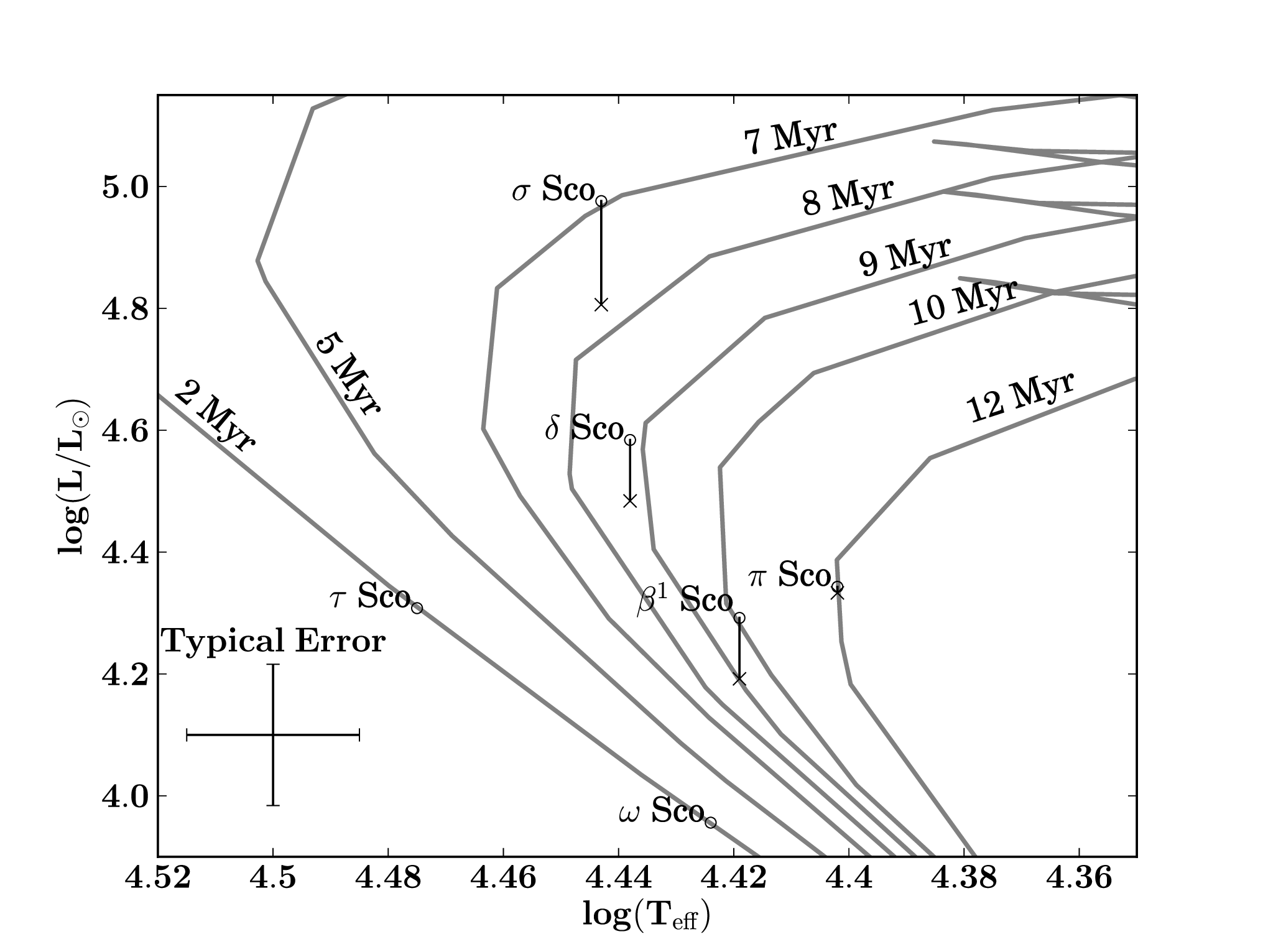}
\caption{\label{fig:usturnoff}
  Upper Sco main sequence turn-off plotted with the
   \citet{1994A&AS..106..275B} evolutionary tracks.  The crosses represent
  the H-R diagram position corrected for known binarity (see text for a discussion
  of each binary).  The circles are the H-R diagram positions uncorrected for 
  companions.  $\omega$~Sco and $\tau$~Sco do not have known companions.
}
\end{center}
\end{figure}

\begin{figure}
\begin{center}
\includegraphics[scale=0.45]{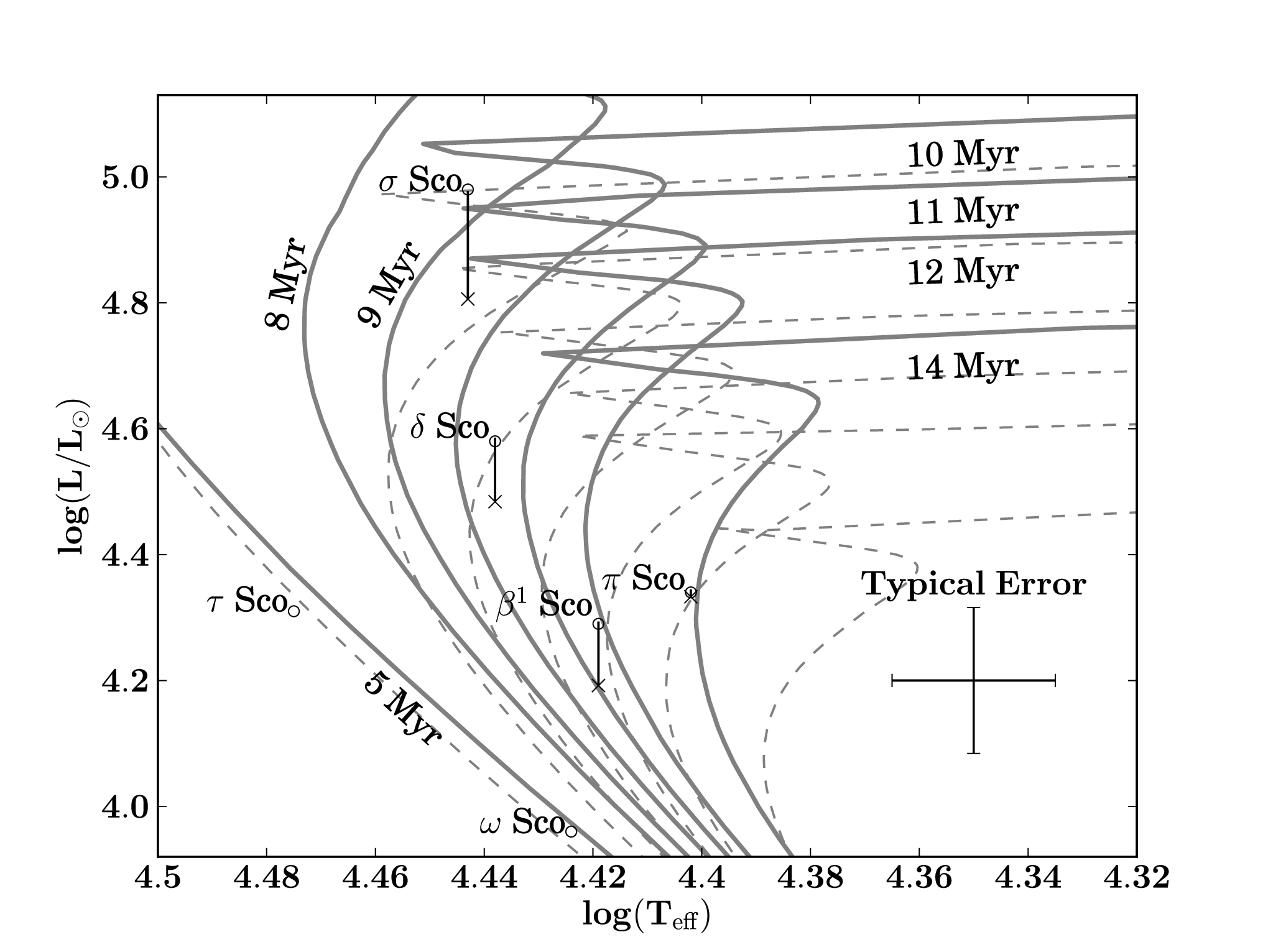}
\caption{\label{fig:usturnoff_rot}
  Upper Sco main sequence turn-off plotted with the
   \citet{2011arXiv1110.5049E} evolutionary tracks without rotation (dashed lines)
   and with v=0.4v$_{\rm breakup}$ (solid lines).  The crosses and circles 
   represent the H-R diagram positions corrected and uncorrected
   for known binarity, respectively.
}
\end{center}
\end{figure}

Unfortunately, the massive stars in Upper Scorpius do not trace a well-defined turn-off, 
even with corrections for known companions. Including the shifted H-R diagram positions 
when accounting for the binarity, we estimate the turnoff age from the following stars:
$\omega$~Sco (2~Myr), $\tau$~Sco (2~Myr), $\beta^1$~Sco (9~Myr), $\delta$~Sco (9~Myr), 
$\pi$~Sco (12~Myr), and $\sigma$~Sco (8~Myr).  
The median age for these stars is 9$\pm$3~Myr (68\% C.L.) with 1$\sigma$ dispersion 4~Myr, midway
between the previous age estimates of $\sim$5~Myr and our value of $\sim$13~Myr from the 
F-type members.

While most current stellar evolutionary models have ignored rotation,
theoretical studies have shown that a moderate rotational velocity of
$v_{eq}\sim$200~km~s$^{-1}$ will increase main sequence lifetimes about 20--30\%.
\citep{2000A&A...361..101M,2000ARA&A..38..143M,1997A&A...322..209T}.
This represents a significant difference in ages obtained with evolutionary 
tracks ignoring rotation vs. those which include a treatment of rotation.
For this reason, we also considered the evolutionary tracks of \citet{2011arXiv1110.5049E}
which include tracks for stars with and without rotation.  It should be noted
that neither the \cite{1994A&AS..106..275B} models nor the \cite{2011arXiv1110.5049E} 
models considered here include pre-main sequence evolution, but since the 
pre-main sequence evolution for a $\sim$9~\msun\, star is $\sim$10$^5$~yr 
\citep{1965ApJ...141..993I} the pre-main sequence time can be safely neglected. Plotted in 
Figure~\ref{fig:usturnoff_rot} is the H-R diagram positions of the US main 
sequence turnoff stars along with isochrones generated from the evolutionary tracks
of \citet{2011arXiv1110.5049E} with no rotation (dashed lines) and those with a rotational 
velocity equal to 40\% of the breakup velocity.  The median age with rotation at 40\%
of breakup is 10$\pm$2~Myr (68\% C.L.) while the median age without is 9$\pm$2~Myr (68\% C.L.).
Since the US turnoff stars are rotating with a median projected rotational velocity of 
$<$\vsini$>$=96 km~s$^{-1}$ (see Table~\ref{tbl:usturnoff}) and $<v_{eq}>$=$\frac{4}{\pi}$$<$\vsini$>$, 
then the turnoff stars have $<v_{eq}>\simeq$120~km~s$^{-1}$.  Using mass estimates from the 
\citet{1994A&AS..106..275B} tracks, we estimate the median break-up velocity of the US turnoff
stars as $\sim$450 km~s$^{-1}$ and thus the median observed rotational velocity is $\sim$25\%
of breakup.  This is a significant median rotational velocity and indicates that rotation 
should be considered.  Therefore we adopt the median age obtained with the rotating 
evolutionary tracks of 10$\pm$2 Myr (68\% C.L.).  

Our Upper Sco turn-off age is much older than the age derived by the \citet{1989A&A...216...44D} study.  
We believe the major source of discrepancy between the \citet{1989A&A...216...44D} US turn-off ages
and our US turn-off ages are the updated evolutionary tracks.  If we use the \citet{1989A&A...216...44D}
H-R diagram positions with the \citet{1994A&AS..106..275B} evolutionary tracks, we obtain a median
age of 9~Myr, the same age we obtain with our modern data (though our individual H-R diagram 
positions are not the same as those in the \citealt{1989A&A...216...44D} study).  We also note that
the use of evolutionary tracks which account for rotation tends to increase the derived ages by 
$\sim$25\% in general \citep{2000A&A...361..101M}.  For example, the using the \citet{2011arXiv1110.5049E}
rotating evolutionary tracks for a typical upper main sequence star in LCC and UCL gives an age 
$\sim$30\% older than ages derived with the \citet{1994A&AS..106..275B} evolutionary tracks.
Revisiting the turn-off ages for UCL and LCC is beyond the scope of this paper, but may be considered
in a future paper.

\subsection{Antares\label{sct:antares}}
\begin{figure}
\begin{center}
\includegraphics[scale=0.45]{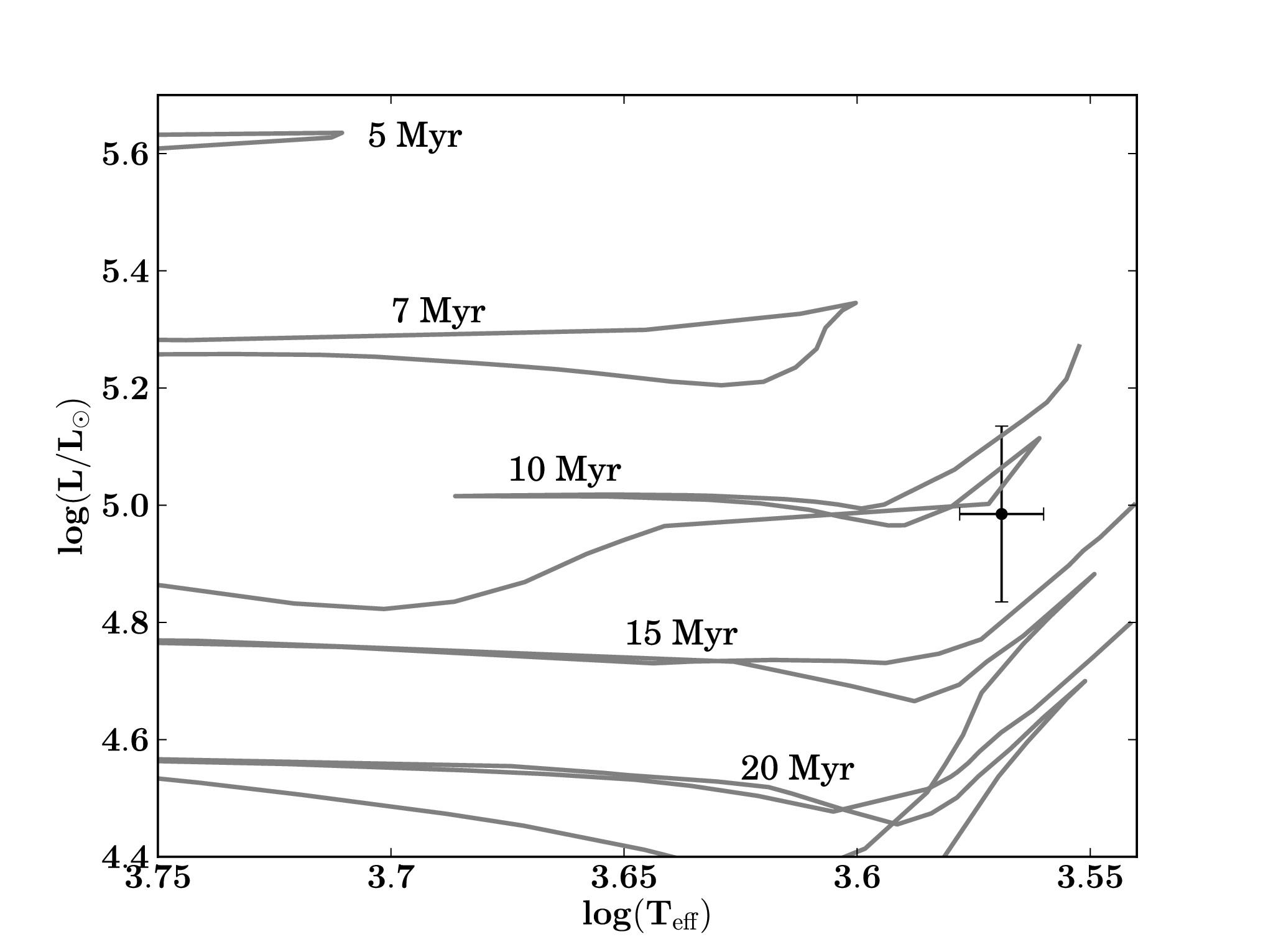}
\caption{\label{fig:antares}
  The H-R diagram for M1.5Iab-Ib supergiant Antares ($\alpha$~Sco).  The
  theoretical isochrones overlap in this region, but the best fit is 11$^{+3}_{-1}$~Myr
  using \citet{1994A&AS..106..275B} evolutionary tracks.
}
\end{center}
\end{figure}

\begin{figure}
\begin{center}
\includegraphics[scale=0.45]{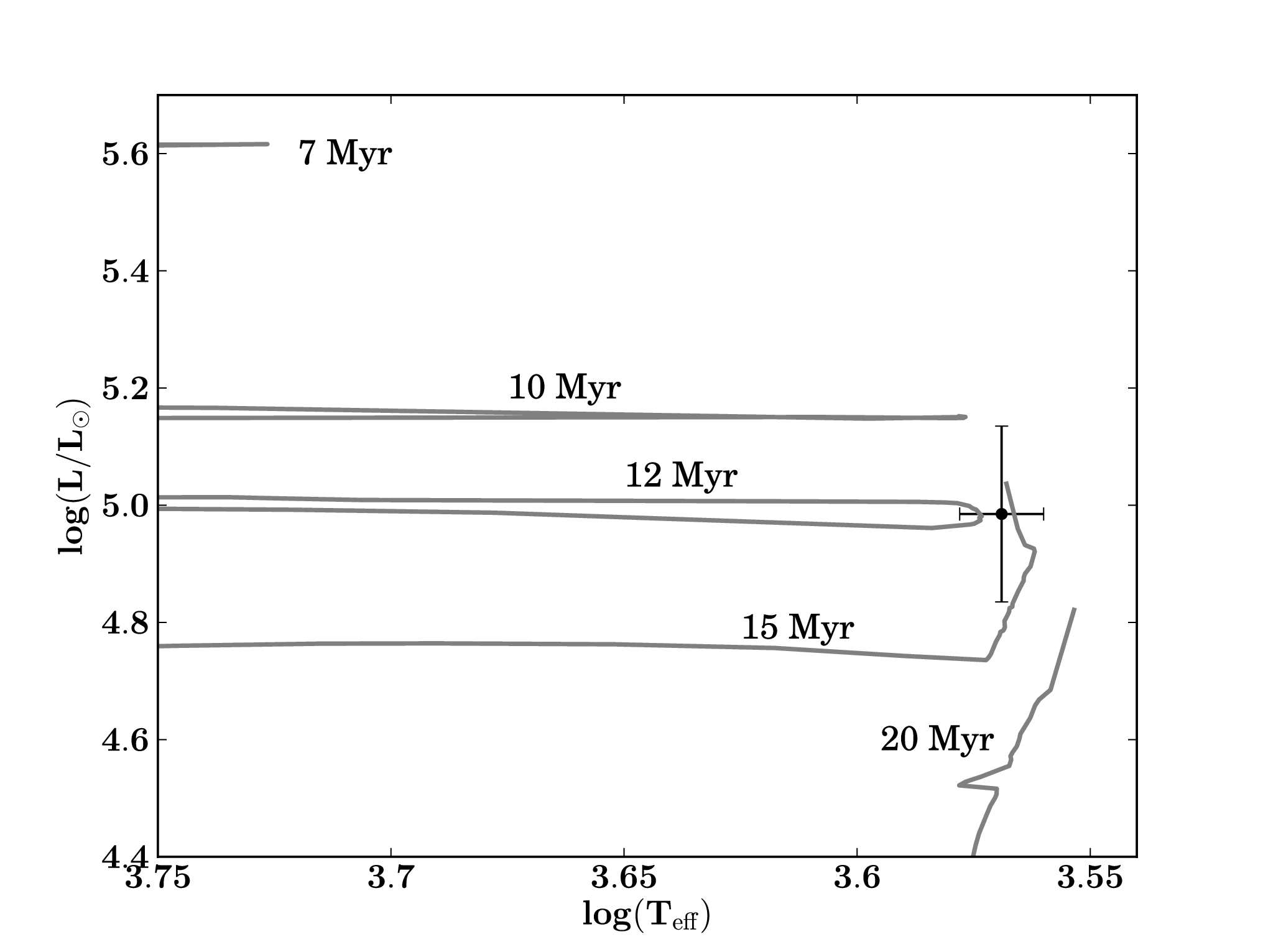}
\caption{\label{fig:antares_rot}
  The H-R diagram for Antares with the rotating evolutionary tracks of 
  \citet{2011arXiv1110.5049E}.  We adopt a final age of Antares of 12$^{+3}_{-1}$~Myr.
}
\end{center}
\end{figure}

Antares ($\alpha$~Sco) is a rare M1.5Iab-Ib supergiant \citep{1989ApJS...71..245K} in Upper Scorpius.
As the most massive secure member of US, it places tight constraints on the age of 
the group.   To estimate the H-R diagram of Antares we use the updated temperature scale
of \citet{2005ApJ...628..973L} for red supergiants, together with the revised {\it Hipparcos} 
parallax \citep{2007A&A...474..653V} of $\pi$=5.89$\pm$1.00~mas and the angular diameter 
measurement of 41.3$\pm$1.0~mas \citep{1990A&A...230..355R}.  We obtain 
log(\teff)=3.569$\pm$0.009 dex (assuming spectral type uncertainty of 1 subtype), 
log(L/\lsun)=4.99$\pm$0.15 dex.  This H-R diagram position is plotted in 
Figure~\ref{fig:antares}.  With this H-R diagram position we obtain an initial mass of
16.6~\msun\, and an age of 11$^{+3}_{-1}$~Myr  with the 
evolutionary tracks of \citet{1994A&AS..106..275B}.
However, just as with the main sequence turn-off stars, the main sequence lifetime of
Antares was likely significantly altered by rotation and therefore we again consider the
rotating evolutionary tracks of \citet{2011arXiv1110.5049E}, shown in Figure~\ref{fig:antares_rot}.
From these rotating evolutionary tracks we obtain an initial mass of 17.2~\msun\, and an 
age of 12$^{+3}_{-1}$~Myr, which we adopt as our final age for Antares.

However, independent of this H-R diagram position we can still place constraints on the age of 
Antares using only the \teff\, derived from the spectral type.  Using the previously mentioned
\teff\, calibration for supergiants, a \teff\, of 3710~K (log(\teff)=3.569) means that 
the age must be 9~Myr or more, since younger isochrones are not predicted to reach 
temperatures that low. 
This is illustrated in Figure~\ref{fig:antares} with the placement of the 5~Myr
isochrone to the left.

\subsection{Ages of Upper Sco A-Type Stars}
We place the A-type stars of US on the H-R diagram as another independent age indicator.  
The F-type stars do not appear to be reaching the main sequence and thus we expect that some
A-type stars will be pre-main sequence and therefore useful as an age indicator.

\begin{figure}
\begin{center}
\includegraphics[scale=0.45]{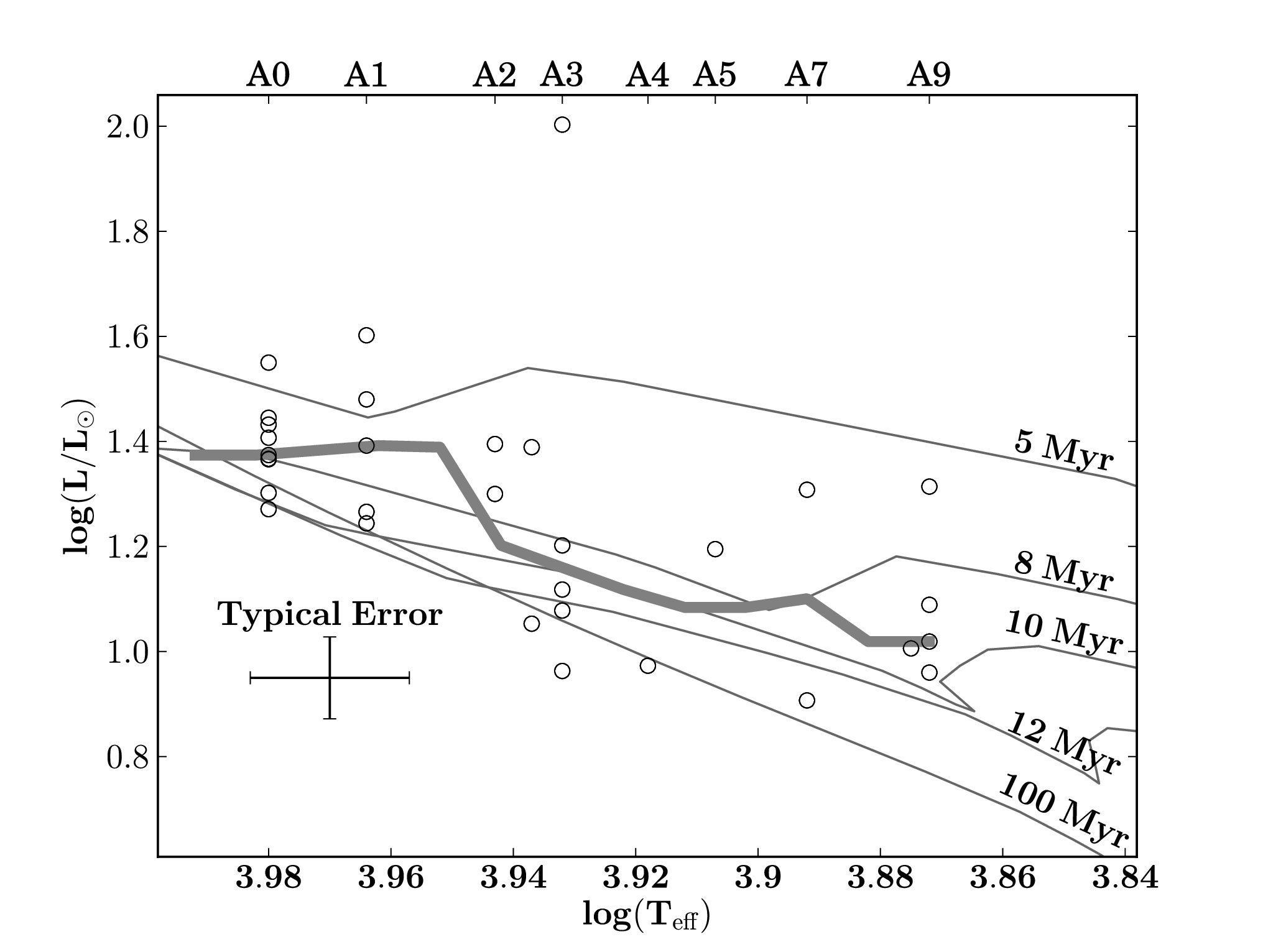}
\caption{\label{fig:hrd_astars}
  H-R diagram for Upper Sco A-type stars taken from \citet{1999AJ....117..354D}.
  Plotted for comparison are 5, 8, 10, 12 and 100 Myr isochrones (solid lines)
  from the \citet{2008ApJS..178...89D} models.
  The star well above the 5~Myr isochrone is HD~150193 (HIP~81624), a known
  Herbig Ae/Be star \citep{2005AJ....129..856H}.  While situated in de Zeeuw et al's Upper Sco 
  ``box'', it appears to be associated with the Oph filamentary clouds L1729 and 1712,
  far from most of the other Upper Sco members.   The thick solid line is an 
  empirical isochrone constructed by taking the median luminosity over a 0.01 dex \teff\,
  step size with a 0.035 dex window.  Uncertainties in log(\teff) are
  determined assuming a spectral type uncertainty of 1 subtype.
}
\end{center}
\end{figure}

We use the kinematically-selected A-type US members of \cite{1999AJ....117..354D} and perform 
essentially the same analysis as used on the F-type stars.  We estimate individual reddenings
using the procedure described in Section~\ref{sct:extinction}.  We calculate a kinematic
parallax, again using the revised {\it Hipparcos} astrometry if the solution is that 
of a single star and use the Tycho-2 proper motions otherwise.  We reject one star, HIP~77457, 
since its trigonometric parallax of 8.58$\pm$0.86 mas disagrees with the kinematic parallax of 
5.10$\pm$0.42 mas by more than 3$\sigma$.  
Our individual H-R diagram position data for the US A-type stars is listed in 
Table~\ref{tbl:atype_params} and plotted in Figure~\ref{fig:hrd_astars}.  As with the F-type 
stars, we do not see the A-type stars clustered around the 5~Myr isochrone, but rather mostly 
between the 5~Myr and 12~Myr isochrones.

While the A-type stars in US have individual H-R diagram positions that are too close to 
the main sequence to yield reliable ages, we can examine the trend of the empirical isochrone
and determine what ages are consistent with the ensemble.  Shown in Figure~\ref{fig:hrd_astars}
is the empirical isochrone, which we use to estimate the main sequence turn-on for Upper Sco
at spectral type $\sim$A3 or log(\teff)$\simeq$3.93~dex.  We compare our empirical isochrones
with the \citet{2008ApJS..178...89D} evolutionary models, noting that an age of 8~Myr is too
young to be consistent with the clump of stars at spectral type A3, and an age of 12~Myr is 
too old to be consistent with the clump of stars at spectral type A8-A9.  Using these 
constraints, we estimate an age of 10$\pm$2~Myr with the \citet{2008ApJS..178...89D} 
evolutionary tracks.  We obtain similar age estimates with other evolutionary models, 
summarized in Table~\ref{tbl:astar_ages}.  For the A-type stars we adopt the median age among 
all four evolutionary models considered of 10$\pm$1$\pm$1~Myr (statistical, systematic).


\subsection{Ages of Upper Sco Pre-Main Sequence G-type Stars Revisited}
Since our F-type median ages for LCC and UCL agree very well with previous results obtained with the 
G-type pre-main sequence stars \citep{2002AJ....124.1670M}, we wish to directly compare
our new results for Upper Scorpius with a similar sample of G-type stars in an attempt
to resolve this discrepancy in ages.  We use the X-ray selected G-type stars from \citet{1999AJ....117.2381P}
and the kinematically selected G-type stars from \citet{1999AJ....117..354D}.
Ages for the X-ray selected objects from the \citet{1999AJ....117.2381P} study had been estimated previously 
using UKST Schmidt plate photometry. However, more precise Tycho-2 and 2MASS photometry are 
now available and thus we may be able to obtain a more reliable age estimate.  Ages for the US G-type
stars in the \citet{1999AJ....117..354D} sample have not been previously examined.

\begin{figure}
\begin{center}
\includegraphics[scale=0.45]{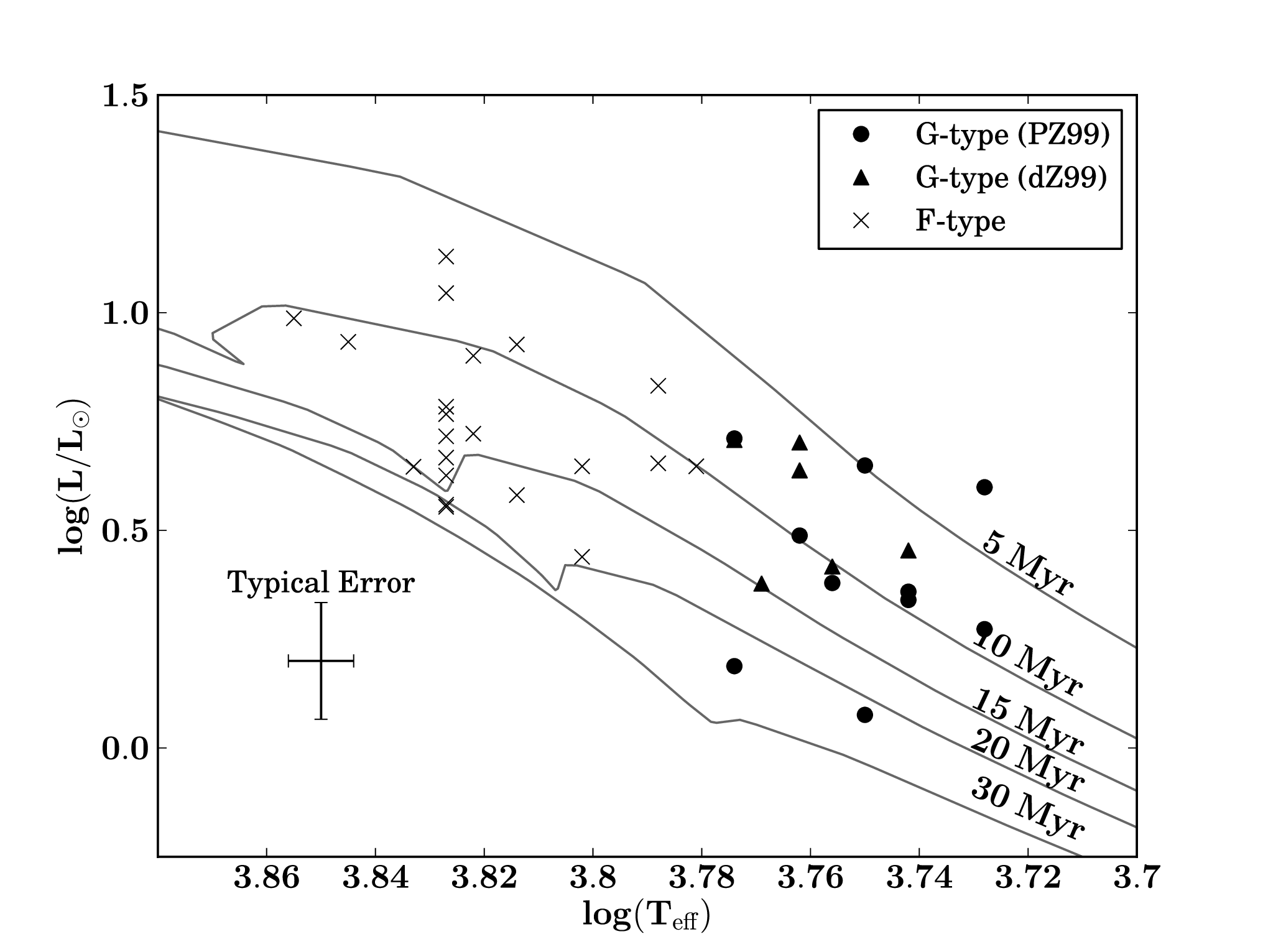}
\caption{\label{fig:pz99_gk}
  H-R diagram for Upper Sco G-type stars taken from \citet{1999AJ....117.2381P} (solid triangles) and 
  \citet{1999AJ....117..354D} (solid circles) with Tycho-2 proper motions within the 
  \cite{1999AJ....117..354D} boundaries.  F-type members (this work) are shown for 
  continuity (crosses).  Plotted for comparison are 5, 10, 15, 20 and 30 Myr isochrones
  from the \citet{2008ApJS..178...89D} models.
}
\end{center}
\end{figure}

For the X-ray-selected sample we cross-reference the \cite{1999AJ....117.2381P} sample 
with the Tycho-2 catalog \citep{2000A&A...355L..27H} and select objects which have proper motions within 
the \cite{1999AJ....117..354D} proper motion boundaries for Upper Sco (i.e., $0<\mu<37$ mas~yr$^{-1}$).   
This leaves us with ten X-ray selected G-type US members.
We correct for extinction using (B-V), (V-J), (V-H), and (V-K) colors from Tycho-2 and 2MASS photometry, 
with the Tycho-2 photometry converted to Johnson using the conversions found in 
\citet{2002AJ....124.1670M,2006AJ....131.2360M}.  To estimate bolometric luminosity we use the converted V-band 
photometry, kinematic distances using the Tycho-2 proper motions and temperatures and bolometric corrections 
listed in Table~\ref{tbl:ic_teff}. 
For the kinematically-selected sample, we first identify two interlopers, HIP~83542 and HIP~81392.  
\citet{1988mcts.book.....H} classified HIP~83542 as G8/K0 III, a giant, and the H-R diagram position is 
well above the 1~Myr isochrones, consistent with this assessment.  HIP~81392 and HIP~83542 both exhibit 
weak lithium absorption features, significantly lower than the other six kinematically-selected US 
candidates from \citet{1999AJ....117..354D} and inconsistent with the youth of Upper Sco (E. Mamajek, private
communication 2011).  With these interlopers removed, we are then left with six G-type US members from 
\citet{1999AJ....117..354D}.  For the \citet{1999AJ....117..354D} G-type US candidate members we 
estimate extinction, {\teff\,}, distance, and bolometric luminosity using the same method as for the F-type 
stars as previously discussed.  Our newly-derived extinctions, distances, and H-R data is listed in 
Table~\ref{tbl:pz99_gk}, with the H-R diagram for this analysis shown in Figure~\ref{fig:pz99_gk}.

For the 16 G-type members of Upper Sco described above, we find median ages of 8-14~Myr using 
the evolutionary tracks of \citet{2008ApJS..178...89D}, \citet{2004ApJS..155..667D}, 
\citet{2000A&A...358..593S}, and \citet{1997MmSAI..68..807D}.  
Our derived ages for these G-type stars of Upper Sco are summarized in Table~\ref{tbl:us_gstar_ages},
with an overall median age among all tracks of 9$\pm$2$\pm$3~Myr (statistical, systematic).
The kinematically-selected members are slightly more biased towards younger-ages than the 
X-ray selected members.  However, this slight difference may be attributed to the brightness
limits of the {\it Hipparcos} catalog.  The difference may not be significant, though, with 
such small numbers of members.

\subsection{Upper Sco Expansion Age}
One of the earlier published ages for US is an expansion age of $\sim$5~Myr derived by 
\citet{1978ppeu.book..101B}.  Deriving ages using kinematic data holds obvious appeal 
since it does not depend on stellar evolution models.  
However, \citet{1997MNRAS.285..479B}, performing simulations of expanding OB associations and 
analyzing the resulting simulated proper motion data, found major problems with kinematic expansion 
ages which only rely on proper motion data.  They found that expansion ages determined by tracing
the proper motion of the stars back to their smallest configuration, as was performed in 
\citet{1978ppeu.book..101B} to obtain the often quoted age of 5~Myr for US,  always leads to 
underestimated ages and that all age estimates converge to $\sim$4~Myr.  Furthermore, they found 
that the initial size of the OB association provided by this method is always overestimated.  The 
\citet{1997MNRAS.285..479B} study found that kinematic expansion ages determined using proper motion 
data alone are essentially meaningless. 

Radial velocities must be considered in order to derive 
a meaningful expansion age.  Given the availability of good quality trigonometric parallax and 
radial velocity data and the findings of \citet{1997MNRAS.285..479B}, we revisit the expansion 
age for US here.  Following \citet{2005ApJ...634.1385M}, we adopt a Blaauw expansion model 
\citep{1956ApJ...123..408B,1964IAUS...20...50B} which gives the observed radial velocity for 
parallel motion as 

\begin{equation*} 
v_{obs} = v'cos{\lambda} + {\kappa}r + K
\end{equation*}

Where $v'$ is the centroid speed of the group, $\lambda$ is the angular distance to the convergent
point, $v'cos\lambda$ = $v_{\rm pred}$ is the predicted radial velocity with no expansion, 
$\kappa$ is the expansion term in units of km~s$^{-1}$~pc$^{-1}$, $r$ is the distance to the star in pc, 
and $K$ is an offset term which may include intrinsic effects (e.g., convective blueshift or 
gravitational redshift).  We then write the difference between the predicted and observed
radial velocity as 

\begin{equation*} 
v_{obs} - v_{pred} = {\Delta}V_R= {\kappa}r + K
\end{equation*}

so that $\kappa$ is the slope in the plot of ${\Delta}V_R$ versus $r$:

\begin{equation*} 
\kappa = \frac{d({\Delta}V_R)}{dr}
\end{equation*}

The expansion age in Myr is then 

\begin{equation*} 
\tau = \gamma^{-1}\kappa^{-1}
\end{equation*}

where $\gamma$=1.0227~pc~s~Myr$^{-1}$~km$^{-1}$, a conversion factor.

\begin{figure}
\begin{center}
\includegraphics[scale=0.45]{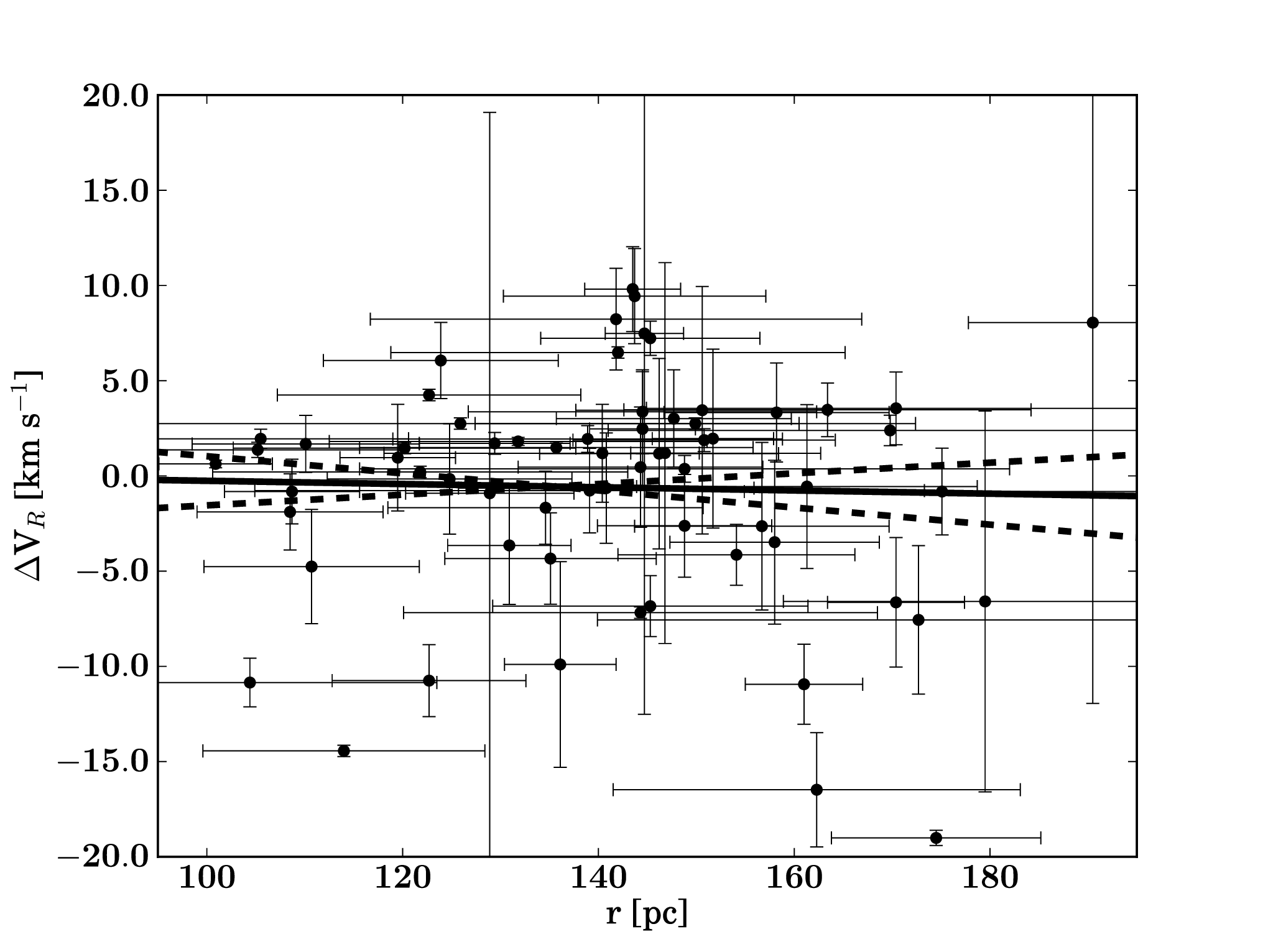}
\caption{\label{fig:us_expand}
  Distance (r) vs. the difference between the observed and predicted radial 
  velocity (V$_R$). The sample includes US candidate members from \citet{1999AJ....117..354D} 
  within 50~pc of the mean US distance and with measured radial velocities 
  within 20~km~s$^{-1}$ of the predicted value.
  If US is expanding, the closest members should be more blueshifted and 
  the further members should be more redshifted.  A series of Monte Carlo
  simulations gives a best-fit slope of $\kappa$ = $-$0.01$\pm$0.04~km~s$^{-1}$~pc$^{-1}$, 
  which places a 99\% confidence lower limit on the expansion age of 10.5~Myr.
}
\end{center}
\end{figure}

To determine the expansion age we have started with the entire \citet{1999AJ....117..354D} sample 
for which radial velocity measurements were available in the literature (listed in 
Table~\ref{tbl:us_expand}), since these all have individual trigonometric parallaxes.  We have
used radial velocities from \citet{2011arXiv1110.0536D}, \citet{2011ApJ...738..122C},
\citet{2006AstL...32..759G}, and \citet{1995A&AS..114..269D}.  We limited the sample to those 
\citet{1999AJ....117..354D} candidate members which have revised {\it Hipparcos} distances 
\citep{2007A&A...474..653V}\footnote{Distances are simply the inverse of the trigonometric parallax.}
within 50~pc of
the mean US distance (145~pc, \citealt{1999AJ....117..354D}).  We also excluded candidate members
with radial velocities more than 20~km~s$^{-1}$ discrepant from the predicted radial velocities, 
as these are almost certainly non-members or unresolved binaries.  Our $v_{\rm pred}$ is calculated using
the new best estimate of the mean UVW motion of Upper Sco, detailed in \citet{2011ApJ...738..122C}. 
The 1-D velocity dispersion in US is 1.3~km~s$^{-1}$ \citep{1999MNRAS.310..585D}.
In order to determine if there is evidence of expansion, we plot the distance $r$ versus ${\Delta}V_R$ 
in Figure~\ref{fig:us_expand}.  Expansion would be exhibited by a 
positive correlation between the distance and ${\Delta}V_R$, characterized by 
the slope ($\kappa$) of a line fit to the data.  To take into account the observational errors in 
distance and radial velocity in fitting a line to the data, we performed a Monte Carlo simulation in 
which we added random Gaussian errors commensurate with each observational error to the observed data 
point, and then we fit a line using an unweighted total least squares algorithm.  We performed 50,000 
trials and then used the median and 1$\sigma$ spread to quantify the effect of the uncertainties in 
the observational data on the slope.
This yielded a Pearson--r of $-$0.03$\pm$0.11 and a best-fit line with a slope of 
$\kappa = -0.01\pm0.04~$km~s$^{-1}$~pc$^{-1}$.  If we consider only models of expansion (i.e., $\kappa>0$),
then the 99\% confidence lower limit on the expansion age is 10.5~Myr.  However, the {\it results are 
statistically consistent with no expansion}.  For an expansion age of 5$\pm$2~Myr we would expect
$\kappa=0.20^{+0.13}_{-0.06}~$km~s$^{-1}$~pc$^{-1}$, and an expansion age of 10$\pm$2~Myr would have 
$\kappa=0.10\pm0.02~$km~s$^{-1}$~pc$^{-1}$.  Given that we are only able to obtain a lower limit 
on the expansion age for US and our expansion data are statistically consistent with no expansion, 
we do not consider our 99\% confidence lower limit in the final age determination for US.  

\subsection{What is the best median age for Upper Sco?}
Every age indicator we have examined thus far has yielded an age of Upper Sco of 9-13 Myr,
summarized in Table~\ref{tbl:us_ages}.
The youngest median age comes from the G-type stars at $\sim$9~Myr, 
and the oldest from the F-type stars at $\sim$13~Myr, so we can conclude with confidence that US 
must be between 9 and 13 Myr old.  Regarding each segment of the H-R diagram as independent, 
we simply take the mean age of our values -- $\sim$10~Myr for the main sequence
turn-off, $\sim$12~Myr for Antares, $\sim$10~Myr for the A-type stars, $\sim$13~Myr for the
F-type members and $\sim$9~Myr for the G-type members.  This gives a final adopted age of
11$\pm$1~Myr (s.e.m., statistical) for Upper Sco, which is more than twice the currently
accepted age \citep{2002AJ....124..404P}.  As each segment of the H-R diagram provides 
an independent age, we estimate our systematic uncertainty as the 1$\sigma$ dispersion of 
these independent ages, which yields a systematic uncertainty of $\pm$2~Myr.

\subsection{Implications of an Older Upper Sco}
Upper Sco has been a benchmark stellar population in studies of circumstellar disk 
lifetimes \citep[e.g.,][]{2006ApJ...651L..49C}, but has a lower primordial disk fraction when 
compared to other stellar populations at 5~Myr (e.g., $\lambda$ Ori; \citealt{2007ApJ...664..481B}).  
Considered in the context of 11~Myr age, the observed primordial disk fraction is consistent with 
those of similarly-aged stellar populations (e.g., NGC~7160, \citealt{2006ApJ...638..897S}).  Our 
detailed analysis of the isochronal ages for massive and intermediate-mass Upper Sco stars may be 
indicative that the ages of similarly aged groups may also be due for further investigation and 
significant revision. \citet{2009MNRAS.399..432N} have also found evidence that the nuclear ages 
for massive stars in $<$10 Myr-old groups are typically 1.5--2$\times$ the pre-MS contraction ages. 
In particular, Naylor (2009) finds that two of Upper Sco's $\sim$5-Myr-old siblings (NGC~2362 
and Cep~OB3b, each with pre-MS ages of 4.5 Myr) have best fit isochronal ages for the massive stars 
of 9.1~Myr and 10~Myr, respectively.  In this regard the doubling of Upper Sco's age does not seem 
surprising.  \citet{2009MNRAS.399..432N} have proposed that the nuclear ages may be the more correct 
ones, and that protoplanetary disks may correspondingly have nearly twice as long to form gas giant 
planets as previously believed.

In addition to circumstellar 
disk studies, Upper Sco has been the subject of many surveys for low-mass companions 
\citep[e.g.,][]{2008ApJ...689L.153L,2011ApJ...726..113I}.  A low-mass companion to 
[PZ99]~J160930.3-210459\footnote{
Although the star is commonly called 1RXS~J160929.1-210524
or 1RXS~1609, this is technically not an appropriate
name. 1RXS is one of the acronyms used for ROSAT X-ray sources,
however it refers to the X-ray source itself rather than any optical
counterpart, as sometimes there may be multiple plausible optical
counterparts (hence why one does not normally encounter stars called
by 1RXS names in the literature, but by ``RX~J'' names or others).  The 
likely optical counterpart (star) was first identified as a pre-MS Upper 
Sco member in \citet{1998A&A...333..619P} (listed as GSC~6213-1358), and
later was referred to as ``Upper Sco 160930.3-210459'' in 
\citet{1999AJ....117.2381P}. SIMBAD standardized the ``Upper Sco''
names to ``[PZ99]~J'' to comply with IAU guidelines for position-based
names, and the [PZ99]~J-names are now in common use.}
was discovered at a very large separation by \citet{2008ApJ...689L.153L}.  The 
mass\footnote{This companion was considered the first directly
imaged exoplanet orbiting a Sun-like star. The discoverers considered
the companion a ``planet'' \citep{2010ApJ...719..497L}
based on their calculated mass and demonstration of common
motion, following the IAU Working Group on Extrasolar Planets (WGESP)
definition, which defined a planet as ``objects with true masses below
the limiting mass for thermonuclear fusion of deuterium (currently
calculated to be 13 Jupiter masses for objects of solar metallicity)
that orbit stars or stellar remnants''; see
http://www.dtm.ciw.edu/boss/definition.html and 
\citealt{2007IAUTA..26..183B}.  
Recent work by \citet{2011ApJ...727...57S} shows that objects
with protosolar composition and masses of $>$11.9 M$_\mathrm{Jup}$ can burn
at least 10\% of their initial deuterium over 10 Gyr.  Hence, following
the definition of ``planet'' and ``brown dwarf'' adopted by the IAU and
the discoverers, the companion should probably be best considered a
``brown dwarf''. Given that the GSC or [PZ99] designation should be
preferred over the 1RXS name, and brown dwarf companion
should probably be called ``GSC~06213-01358~B'' or 
``[PZ99]~J160930.3-210459~B''.} 
of this 
companion was determined to be 8~M$_\mathrm{Jup}$ using the assumed age of 5~Myr.  However, we use our 
revised age for Upper Sco of 11~Myr and the object's reported \teff\, of 1800$\pm$200~K 
\citep{2010ApJ...719..497L} to obtain a mass of 13$^{+2}_{-3}$~M$_\mathrm{Jup}$ using the DUSTY models 
\citep{2000ApJ...542..464C,2002A&A...382..563B}.  Alternatively, using the object's bolometric luminosity 
of log(L/\lsun)=-3.55$\pm$0.2 \citep{2010ApJ...719..497L} and the 11~Myr age we estimate a mass of 
14$^{+2}_{-3}$~M$_\mathrm{Jup}$, consistent with the estimates from \teff\,.  Here we follow 
\citet{2010ApJ...719..497L} and adopt the estimates from the luminosity, listed in 
Table~\ref{tbl:substellar}. 
Similarly, a companion to GSC~06214-00210
was discovered by \citet{2011ApJ...726..113I} and thought to have a mass of $\sim$14~M$_\mathrm{Jup}$ at 
an age of 5~Myr.  However, at an age of 11~Myr the mass is slightly higher at 17$\pm$3~M$_\mathrm{Jup}$, using 
the DUSTY models and the estimated absolute magnitudes M$_J$ $\sim$10.5, M$_H$ $\sim$9.6, M$_K$ $\sim$ 9.1 
\citep{2011ApJ...726..113I}.
The substellar companion to HIP~78530 \citep{2011ApJ...730...42L} was estimated to have a mass of
$\sim$21--26~M$_\mathrm{Jup}$ with an assumed age of 5$\pm$1~Myr.  We use the bolometric luminosity of
log(L/\lsun) = -2.55$\pm$0.13 \citep{2011ApJ...730...42L} and our revised age of 11~Myr to obtain a mass 
for HIP~78530B of 30$^{+17}_{-8}$~M$_\mathrm{Jup}$.
The substellar companion to the young brown dwarf UScoCTIO~108 has a mass of 14$^{+2}_{-8}$~M$_\mathrm{Jup}$ 
with an assumed age of 5--6~Myr \citep{2008ApJ...673L.185B}.  Using the reported luminosity of 
log(L/\lsun)=-3.14$\pm$0.20 \citep{2008ApJ...673L.185B} and our revised age of 11~Myr, we estimate a mass 
of 16$\pm$2~M$_\mathrm{Jup}$ for UScoCTIO~108b.  The substellar binary Oph~J1622-2405 initially 
had component mass estimates of $\sim$14~M$_\mathrm{Jup}$ and $\sim$7~M$_\mathrm{Jup}$ based on an assumed 
age of 1~Myr \citep{2006Sci...313.1279J}.  However, followup spectroscopic studies 
\citep{2007ApJ...657..511A,2007ApJ...660.1492C,2007ApJ...659.1629L} 
found higher masses.  Here we estimate the masses using the effective temperatures from 
\citet{2007ApJ...659.1629L} with our revised age of 11~Myr and find masses of 53$^{+9}_{-7}$~M$_\mathrm{Jup}$ 
and 21$\pm$3~M$_\mathrm{Jup}$ for Oph~J1622-2405A and Oph~J1622-2405B, respectively, with the DUSTY models.  
This is in agreement with the results from \citet{2007ApJ...659.1629L}, who used H-R diagram 
positions with the DUSTY models for their mass estimates. 
The revised mass estimates discussed above are summarized in Table~\ref{tbl:substellar}. 
Hence, we believe that none of the substellar companions directly imaged in Upper Sco 
have inferred masses below the deuterium-burning limit.

Sco-Cen has been used as an example of triggered star formation, with supernova in UCL/LCC
triggering star formation in US and supernova in US triggering star formation in $\rho$ Oph
\citep{1992A&A...262..258D,1999AJ....117.2381P}.  The UCL shell has an expansion velocity of 
10$\pm$2 km~s$^{-1}$ and radius 110$\pm$10~pc, suggesting
it originated in UCL $\sim$11~Myr ago and passed through US about 4~Myr ago \citep{1992A&A...262..258D}.  However,
given the revised age estimates for US, it does not seem likely that the passage of this superbubble
would have triggered star formation in US since it would have arrived too late.  However, this does not
mean that star formation in US was not triggered by UCL, it simply means that the timescales and
ages are not consistent with the arrival of the UCL shell in US as the triggering event.  Similarly,
the shell around US has an expansion velocity of 10$\pm$2 km~s$^{-1}$ and a radius of 40$\pm$4~pc,
consistent with an age for the shell of $\sim$4~Myr. At 10 km~s$^{-1}$ it would have 
traveled the $\sim$15~pc from US to $\rho$ Oph in about 1.5~Myr.  
With the revised 11~Myr age for US and $\rho$ Oph ages of $\sim$2-3~Myr 
\citep{2011AJ....142..140E}, there is sufficient time for star formation in 
$\rho$ Oph to have been triggered by the US shell.

Another important implication of our result is to drive down the
inferred progenitor mass for the runaway young neutron star 
RX~J1856.5-3754. \citet{2011MNRAS.417..617T} claim that RX~J1856 formed in
Upper Sco and was ejected 0.5 Myr ago. Assuming a progenitor mass of 5
Myr, \citet{2011MNRAS.417..617T} predicted that the progenitor would have had a
mass of $\sim$37-45~\msun\, and main sequence spectral type of
O5-O7. \citet{2009ARA&A..47...63S} reviewed the observational and theoretical
constraints on the supernovae and their progenitors, and suggests that
progenitors with masses of $>$30~\msun\, almost certainly form black
holes, not neutron stars. However, if the progenitor of J1856 was a
10.5 Myr-old star when it exploded 0.5~Myr ago, then the progenitor
was most likely a $\sim$18--20~\msun\, O9-type star (using the
\citealt{2011arXiv1110.5049E} tracks), with the remnant being either a neutron
star or black hole (likely depending on the rotation or whether it was
in an interacting binary; \citealt{2009ARA&A..47...63S}). Indeed, if the empirically
constrained set of supernova outcomes outlined by \citet{2009ARA&A..47...63S} are
correct, and if one adopts the previous age of 5~Myr, then the deceased
high mass stars in Upper Sco should have {\it all} turned into black
holes. This would leave the ``young'' neutron star RX~J1856 with a well
constrained distance and proper motion but without a plausible
birth site. We conclude that the RX~J1856 runaway scenario is more
plausible if Upper Sco is 11~Myr, and the progenitor would have been a
much easier to form $\sim$18--20~\msun\, star rather than a rarer 
$\sim$37--45~\msun\, star.

\section{Conclusions}
We can summarize our conclusions as follows:

\begin{enumerate}
\item The pre-main sequence ``turn-on'' for UCL and LCC occur near spectral type $\sim$F2 and $\sim$F4,
respectively.  The F-type members of US appear to be all pre-main sequence.  Examining the A-type
members of US allows us to estimate the pre-main sequence turn-on for US to be near spectral
type $\sim$A3.

\item The good agreement between the isochronal ages of F-type stars studied here and the 
G- and K-type stars studied in \cite{2002AJ....124.1670M} provides greater confidence in 
those ages, and more firmly establishes that UCL is slightly younger than, or roughly coeval 
with, LCC.  In order of youngest to oldest: Upper Scorpius (US), Upper Centaurus-Lupus (UCL), 
Lower Centaurus-Crux (LCC).

\item The median ages obtained with the pre-main sequence F-type members (those later than ~F5) 
for UCL and LCC agree with previous results for early B-type stars and pre-main sequence
G-type stars. \citep{2002AJ....124.1670M}, with median ages of 16$\pm$1~Myr and 17$\pm$1~Myr, 
respectively.

\item The F-type members have a median isochronal age of 13$\pm$1~Myr (68\% C.L.) for US, which is 
much older than previous results.  Using the approximate pre-main sequence turn-on point 
with the A-type stars allows us to estimate an age of 10$\pm$3~Myr.  
Re-examining the early B-type stars and the G-type pre-main sequence stars in US 
yields median ages of 10$\pm$2~Myr and 9$\pm$2~Myr, respectively.  Age estimates
for the M1.5Iab-Ib supergiant US member Antares give 12$^{+3}_{-1}$~Myr.  Considering these to 
be independent estimates, we obtain an overall mean age for Upper Sco of 
11$\pm$1$\pm$2~Myr (statistical, systematic) with the uncertainty in the mean age 
dominated by systematic differences between the isochronal ages inferred from different 
mass ranges. 

\item We find a 99\% confidence lower limit on the kinematic expansion 
age for Upper Sco of 10.5~Myr using trigonometric parallax and radial velocity data.
However, the astrometry and radial velocities are statistically consistent with 
no expansion.

\item Based on H$\alpha$ emission accretion diagnostics, we estimate a spectroscopic accretion
disk fraction of 0/17 ($<$19\%; 95\% C.L.) in US, consistent with the {\it Spitzer}
results of \cite{2006ApJ...651L..49C} of 0/30 ($<$11\%; 95\% C.L.) for F- and G-type stars.  
In UCL we find 1/41 ($2^{+5}_{-1}$\%; 68\% C.L.) accretors and 
in LCC we find 1/50 ($2^{+4}_{-1}$\%; 68\% C.L.) accretors, which are both consistent with the
spectroscopic accretion disk fraction for the G- and K-type members of UCL and LCC from 
\cite{2002AJ....124.1670M} for G- and K-type stars.

\item The revised age of 11~Myr for Upper Sco may explain the lower than expected disk
fraction compared to 5~Myr groups (e.g., $\lambda$~Ori; \citealt{2007ApJ...664..481B}).
We reevaluate the masses of the known substellar companions in Upper Sco using available 
published data and our revised age of 11~Myr, and find that inferred masses are 
typically $\sim$20--70\% more massive than previously estimated.  All of the imaged
companions thus far appear to be more massive than the deuterium-burning limit and 
so none of the companions are in the planetary-mass regime.
In addition to larger substellar masses, the older age for Upper Sco implies a more
realistic, lower progenitor mass of $\sim$18--20~\msun\, for the young neutron star 
RX~J1856.5-3754.

\end{enumerate}

\acknowledgements
We thank Fred Walter for the use of his SMARTS RC-spectrograph pipeline for reducing the data
obtained at the SMARTS 1.5m telescope at Cerro Tololo, Chile, as well as Jose Velasquez and 
Manuel Hernandez for their help and advice at the telescope.  University of Rochester is a member
of the Small and Moderate Aperture Research Telescope System (SMARTS) Consortium.  This work 
has been supported by NSF grant AST-1008908 and funds from the School of Arts and Sciences at the 
University of Rochester.  \\

\bibliography{ms}

\clearpage

\LongTables
\begin{landscape}
\begin{deluxetable}{llrrrrrrrrrl}
\tabletypesize{\scriptsize}
\tablewidth{0pt}
\tablecaption{Input Observables For Sco-Cen Candidate Members \label{tbl:inputparams}}
\tablehead{
  \colhead{HIP} &
  \colhead{Group} &
  \colhead{2MASS} & 
  \colhead{$\mu_{\alpha*}$} & 
  \colhead{$\mu_{\delta}$} & 
  \colhead{V} &
  \colhead{B-V} &
  \colhead{J} &
  \colhead{H} &
  \colhead{K$_S$} &
  \colhead{PM} \\
  \colhead{} & 
  \colhead{} & 
  \colhead{} & 
  \colhead{(mas~yr$^{-1}$)} & 
  \colhead{(mas~yr$^{-1}$)} & 
  \colhead{(mag)} & 
  \colhead{(mag)} & 
  \colhead{(mag)} & 
  \colhead{(mag)} & 
  \colhead{(mag)} & 
  \colhead{Ref.}
}
\startdata
55334 & LCC & 11195276-7037065 & -43.28$\pm$0.66 & 4.41$\pm$0.58 & 8.14 & 0.413$\pm$0.004 & 7.319$\pm$0.024 & 7.153$\pm$0.027 & 7.084$\pm$0.021 & H \\
56227 & LCC & 11313495-6743545 & -41.49$\pm$0.61 & -3.09$\pm$0.51 & 8.34 & 0.316$\pm$0.017 & 7.779$\pm$0.027 & 7.683$\pm$0.036 & 7.612$\pm$0.021 & H \\
56673 & LCC & 11371464-6940272 & -42.40$\pm$1.30 & -5.20$\pm$1.30 & 6.62 & 0.518$\pm$0.006 & 5.688$\pm$0.024 & 5.459$\pm$0.024 & 5.326$\pm$0.023 & T \\
57595 & LCC & 11482700-5409185 & -27.60$\pm$2.00 & -4.50$\pm$1.90 & 9.45 & 0.475$\pm$0.024 & 8.512$\pm$0.027 & 8.285$\pm$0.047 & 8.215$\pm$0.024 & T \\
57950 & LCC & 11530799-5643381 & -38.41$\pm$0.71 & -10.37$\pm$0.58 & 8.26 & 0.401$\pm$0.016 & 7.467$\pm$0.024 & 7.312$\pm$0.027 & 7.276$\pm$0.027 & H \\
58075 & LCC & 11543559-5443572 & -29.38$\pm$0.83 & -7.05$\pm$0.82 & 9.02 & 0.418$\pm$0.019 & 8.228$\pm$0.056 & 7.997$\pm$0.047 & 7.925$\pm$0.027 & H \\
58146 & LCC & 11552884-6211471 & -34.95$\pm$0.46 & -4.51$\pm$0.39 & 7.86 & 0.391$\pm$0.012 & 7.051$\pm$0.020 & 6.914$\pm$0.024 & 6.833$\pm$0.027 & H \\
58167 & LCC & 11554354-5410506 & -36.17$\pm$0.73 & -8.56$\pm$0.62 & 8.30 & 0.404$\pm$0.013 & 7.473$\pm$0.024 & 7.361$\pm$0.051 & 7.284$\pm$0.020 & H \\
\enddata
\tablecomments{Proper motion references: (H) -- \citet{2007A&A...474..653V}, (T) -- \citet{2000A&A...355L..27H}}
\end{deluxetable}
\clearpage
\end{landscape}

\begin{deluxetable}{lllrlr}
\tabletypesize{\scriptsize}
\tablewidth{0pt}
\tablecaption{Spectral Standard Stars Used for Classification\label{tbl:spstands}}
\tablehead{
  \colhead{Standard} &
  \colhead{Spectral} &
  \colhead{B-V} & 
  \colhead{Instrument/Source} & 
  \colhead{References} \\
  \colhead{} & 
  \colhead{Type} & 
  \colhead{(mag)} & 
  \colhead{} & 
  \colhead{} & 
  \colhead{}}
\startdata
  HD 158352 & A8 V      & 0.237  & SMARTS 1.5m/Rochester          & 1, A \\
  HD 73450  & A9 V      & 0.251  & SMARTS 1.5m/Rochester          & 2, B \\
  HD 23585  & F0 V      & 0.291  & SMARTS 1.5m/NStars             & 3, C \\
  HD 27397  & F0 IV     & 0.283  & SMARTS 1.5m/Stony Brook        & 3, A \\
  HD 89025  & F0 IIIa   & 0.307  & SMARTS 1.5m/Rochester          & 3, A \\
  HD 167858 & F1 V      & 0.312  & SMARTS 1.5m/Rochester          & 4, A \\
  HD 113139 & F2 V      & 0.368  & DSO/NStars                     & 3, A \\
  HD 40535  & F2 III-IV & 0.333  & SMARTS 1.5m/Stony Brook        & 4, A \\
\enddata
\tablecomments{
Notes: Spectral Type sources: 
(1) \citet{1969AJ.....74..375C}; 
(2) \citet{2002AJ....124..989G}; 
(3) \citet{2001AJ....121.2148G}; 
(4) \citet{1989ApJS...69..301G}; 
(5) \citet{1965ApJ...141..177M};
(6) \citet{2006AJ....132..161G};
(7) \citet{1989AJ.....98.1049G};
B-V color sources: 
(A) \citet{2007A&A...474..653V}; 
(B) \citet{1980A&AS...42..251N}; 
(C) \citet{2006yCat.2168....0M};
}
\end{deluxetable}

\clearpage
\begin{deluxetable}{lrrrrrrrr}
\tabletypesize{\scriptsize}
\tablewidth{0pt}
\tablecaption{Intrinsic Colors of Dwarfs and Adopted \teff, BC Values\label{tbl:ic_teff}}
\tablehead{
  \colhead{Spectral Type} & 
  \colhead{U-B} &
  \colhead{B-V} &
  \colhead{V-I$_C$} & 
  \colhead{V-J} & 
  \colhead{V-H} & 
  \colhead{V-K$_S$} & 
  \colhead{BC$_V$} & 
  \colhead{\teff} \\
  \colhead{} &  
  \colhead{(mag)} & 
  \colhead{(mag)} & 
  \colhead{(mag)} & 
  \colhead{(mag)} & 
  \colhead{(mag)} & 
  \colhead{(mag)} & 
  \colhead{(mag)} & 
  \colhead{(K)} }
\startdata
A0V & -0.014 & 0.000 & 0.004 & 0.046 & 0.014 & 0.042 & -0.17 & 9550  \\
A1V & 0.033 & 0.043 & 0.044 & 0.094 & 0.070 & 0.101 & -0.11 & 9200  \\
A2V & 0.063 & 0.074 & 0.092 & 0.167 & 0.158 & 0.192 & -0.05 & 8760  \\
A3V & 0.077 & 0.090 & 0.110 & 0.197 & 0.196 & 0.231 & -0.02 & 8550  \\
A4V & 0.097 & 0.140 & 0.165 & 0.296 & 0.318 & 0.355 & 0.00 & 8270  \\
A5V & 0.100 & 0.160 & 0.187 & 0.334 & 0.366 & 0.404 & 0.01 & 8080  \\
A6V & 0.098 & 0.170 & 0.198 & 0.355 & 0.391 & 0.429 & 0.02 & 8000  \\
A7V & 0.091 & 0.210 & 0.242 & 0.433 & 0.488 & 0.528 & 0.04 & 7800  \\
\enddata
\end{deluxetable}

\clearpage

\LongTables
\begin{landscape}
\begin{deluxetable}{lllrrrrrlrlrlrlrl}
\tabletypesize{\scriptsize}
\setlength{\tabcolsep}{0.04in}
\tablewidth{0pt}
\tablecaption{Stellar Parameters for F-type Sco-Cen Members \label{tbl:stellarparams}}
\tablehead{
  \colhead{HIP} &
  \colhead{Group} &
  \colhead{Spectral} & 
  \colhead{$\pi_{kin}$} &
  \colhead{EW(H$\alpha$)} &
  \colhead{A$_V$} &
  \colhead{log(\teff)} & 
  \colhead{log(L/\lsun)} & 
  \colhead{Mass} & 
  \colhead{Age}  &
  \colhead{Mass} & 
  \colhead{Age}  &
  \colhead{Mass} & 
  \colhead{Age}  &
  \colhead{Mass} & 
  \colhead{Age}  &
  \colhead{Median} \\
  \colhead{} & 
  \colhead{} & 
  \colhead{Type} & 
  \colhead{} & 
  \colhead{} & 
  \colhead{} & 
  \colhead{} & 
  \colhead{} & 
  \colhead{D08} & 
  \colhead{D08} & 
  \colhead{YY04} & 
  \colhead{YY04} & 
  \colhead{S00} & 
  \colhead{S00} & 
  \colhead{DM97} & 
  \colhead{DM97} & 
  \colhead{Mass}  \\
  \colhead{} & 
  \colhead{} & 
  \colhead{} & 
  \colhead{(mas)} & 
  \colhead{(\text{\AA})} & 
  \colhead{(mag)} & 
  \colhead{(dex)} & 
  \colhead{(dex)} & 
  \colhead{(\msun)} & 
  \colhead{(Myr)} & 
  \colhead{(\msun)} & 
  \colhead{(Myr)} & 
  \colhead{(\msun)} & 
  \colhead{(Myr)} & 
  \colhead{(\msun)} & 
  \colhead{(Myr)} &
  \colhead{(\msun)}
}
\startdata
55334 & LCC & F2V & 12.63$\pm$0.90 & 4.7 & 0.12$\pm$0.01 & 3.833$\pm$0.006 & 0.49$\pm$0.06 & 1.5$^\ast$ & $>$19 & 1.5$^\ast$ & $>$22 & 1.6$^\ast$ & ... & 1.5$^\ast$ & $>$25 & 1.5 \\
56673 & LCC & F5IVe & 11.95$\pm$0.90 & -1.9 & 0.20$\pm$0.05 & 3.814$\pm$0.011 & 1.18$\pm$0.07 & 2.0 & 5 & 2.0 & 5 & 2.0 & 7 & 2.0 & 5 & 2.0 \\
57950 & LCC & F3V & 10.79$\pm$0.73 & 4.9 & 0.04$\pm$0.01 & 3.827$\pm$0.005 & 0.55$\pm$0.06 & 1.4 & 23 & 1.4 & 27 & 1.6$^\ast$ & $>$20 & 1.5$^\ast$ & $>$14 & 1.5 \\
58075 & LCC & F2V & 8.13$\pm$0.57 & 5.1 & 0.14$\pm$0.04 & 3.833$\pm$0.006 & 0.53$\pm$0.06 & 1.5$^\ast$ & $>$14 & 1.5$^\ast$ & $>$15 & 1.6$^\ast$ & $>$30 & 1.5$^\ast$ & $>$17 & 1.5 \\
58146 & LCC & F3V & 9.67$\pm$0.65 & 5.1 & 0.04$\pm$0.02 & 3.827$\pm$0.005 & 0.80$\pm$0.06 & 1.5 & 13 & 1.6 & 13 & 1.6 & 13 & 1.6 & 10 & 1.6 \\
58167 & LCC & F2V & 9.96$\pm$0.67 & 5.2 & 0.09$\pm$0.02 & 3.833$\pm$0.006 & 0.62$\pm$0.06 & 1.5 & 19 & 1.5 & 21 & 1.6$^\ast$ & $>$16 & 1.5 & 22 & 1.5 \\
58220 & LCC & F4V & 10.39$\pm$0.71 & 4.3 & 0.08$\pm$0.04 & 3.822$\pm$0.009 & 0.51$\pm$0.06 & 1.4 & 23 & 1.4 & 28 & 1.5$^\ast$ & $>$22 & 1.5$^\ast$ & $>$15 & 1.5 \\
58528 & LCC & F5V & 10.68$\pm$0.72 & 4.3 & 0.04$\pm$0.06 & 3.814$\pm$0.011 & 0.44$\pm$0.06 & 1.4 & 21 & 1.4 & 26 & 1.5$^\ast$ & $>$23 & 1.4$^\ast$ & $>$15 & 1.4 \\
\enddata
\tablecomments{Spectral Types are from this work (Section \ref{sct:spt}).  Uncertainties in Log(\teff) are 
calculated with a spectral type uncertainty of 1 subtype.
In cases where the H-R diagram position was too close to the main sequence to yield a unique age, we
provide upper limits (denoted by $<$) based on the 2$\sigma$ upper limit on the luminosity.  We have provided 
the median mass of all evolutionary tracks as the preferred mass for each member.  Given the
sizes of the observational uncertainties, it is preferable to adopt the median subgroup ages rather 
than these individual isochronal ages.\\
$\ast$ -- mass was derived assuming a 15~Myr age.. $\dagger$ -- mass was derived assuming a 10~Myr age.}
\end{deluxetable}
\clearpage
\end{landscape}

\begin{deluxetable}{llll}
\tabletypesize{\scriptsize}
\tablewidth{0pt}
\tablecaption{Stars Rejected as Sco-Cen Members \label{tbl:rejected}}
\tablehead{
  \colhead{Name} &
  \colhead{Spectral} & 
  \colhead{Rejection Criteria} &
  \colhead{Ref.} \\
  \colhead{} & 
  \colhead{Type} & 
  \colhead{} &
  \colhead{}
}
\startdata
HIP~56227 & F0IV      & Kinematic Parallax    & 1 \\
HIP~57595 & F6V       & Trigonometric Parallax & 1 \\
HIP~59781 & F8V       & Kinematic Parallax    & 1 \\
HIP~61086 & F1V       & HRD Position          & 1 \\
HIP~62056 & F6V       & HRD Position          & 1 \\
HIP~62428 & A7III     & Giant                 & 1 \\
HIP~62674 & F3V       & HRD Position          & 1 \\
HIP~63022 & F0V       & HRD Position          & 1 \\
HIP~64316 & F3V       & HRD Position          & 1 \\
HIP~66285 & F7V       & Companion Li-poor     & 1 \\
HIP~69327 & F1V       & Kinematic Parallax    & 1 \\
HIP~70833 & F3V       & Companion Li-poor     & 1 \\
HIP~75824 & F3V       & Li-poor               & 1 \\
HIP~77457 & A7IV      & Kinematic Parallax    & 2 \\
HIP~79258 & F4V       & Kinematic Parallax    & 1 \\
HIP~81392 & G2/3 V    & Li-poor               & 2 \\
HIP~83542 & G8/K0 III & Giant                 & 3 
\enddata
\tablecomments{Spectral Type References: (1) this work, (2) \citet{1982mcts.book.....H}
(3) \citet{1988mcts.book.....H}
}
\end{deluxetable}

\begin{deluxetable}{lrrr}
\tabletypesize{\scriptsize}
\tablewidth{0pt}
\tablecaption{Median Age Estimates of the PMS F-Type members of US, UCL and LCC\label{tbl:ages}}
\tablehead{
  \colhead{Evolutionary} &
  \colhead{US} & 
  \colhead{UCL} & 
  \colhead{LCC} \\
  \colhead{Tracks} & 
  \colhead{(Myr)} &
  \colhead{(Myr)} &
  \colhead{(Myr)} \\
}
\startdata
 Dartmouth   & 13$\pm$4 & 16$\pm$5  & 18$\pm$7 \\
 Yonsei-Yale & 13$\pm$4 & 15$\pm$5  & 18$\pm$9 \\
 SDF00       & 14$\pm$5 & 16$\pm$9  & 15$\pm$6 \\
 DM97        & 12$\pm$3 & 15$\pm$6  & 15$\pm$5 \\
\hline \\
Median           & 13      & 16      & 17      \\
Statistical Unc. & $\pm$1  & $\pm$1  & $\pm$1  \\
Systematic Unc.  & $\pm$1  & $\pm$1  & $\pm$2
\enddata
\tablecomments{
Uncertainties represent the 1$\sigma$ dispersion in the Chauvenet's-criterion 
clipped ages \citep{2003drea.book.....B} for the members for which we were 
able to determine an age.  The overall uncertainty is the 68\% C.L. 
of the median \citep{2001ApJ...549....1G} (statistical) and the dispersion
in ages (systematic).  
Median ages for UCL and LCC are only defined by members F5 or cooler, as earlier 
members are too close to the main sequence to yield reliable ages. 
}
\end{deluxetable}

\begin{deluxetable}{lrrrrrrrrr}
\tabletypesize{\scriptsize}
\tablewidth{0pt}
\tablecaption{Stellar Parameters for US Turnoff Stars\label{tbl:usturnoff}}
\tablehead{
  \colhead{Name} &
  \colhead{log(\teff)} & 
  \colhead{log(L/\lsun)} &
  \colhead{\vsini} & 
  \colhead{B94 Age} &
  \colhead{B94 Mass} &
  \colhead{E11 Age} & 
  \colhead{E11 Mass} &
  \colhead{E11 Age} & 
  \colhead{E11 Mass} \\
  \colhead{} &
  \colhead{} &
  \colhead{} &
  \colhead{} &
  \colhead{} &
  \colhead{} & 
  \colhead{(no rot.)} &
  \colhead{(no rot.)} & 
  \colhead{(rot.)} & 
  \colhead{(rot.)} \\
  \colhead{} &
  \colhead{(dex)} &
  \colhead{(dex)} &
  \colhead{(km~s$^{-1}$)} &
  \colhead{(Myr)} &
  \colhead{(\msun)} & 
  \colhead{(Myr)} &
  \colhead{(\msun)} & 
  \colhead{(Myr)} & 
  \colhead{(\msun)} 
}
\startdata
 $\omega$~Sco   & 4.424 $\pm$ 0.019 & 3.96 $\pm$ 0.05 & 100$\pm$6  &  2 & 11.4 &  5 & 11.2 &  5 & 11.4 \\
 $\beta^1$~Sco  & 4.419 $\pm$ 0.010 & 4.29 $\pm$ 0.05 &  91$\pm$8  &  9 & 12.2 & 11 & 12.1 & 11 & 12.2 \\
 $\pi$~Sco      & 4.402 $\pm$ 0.007 & 4.34 $\pm$ 0.10 & 100$\pm$15 & 12 & 12.4 & 12 & 12.9 & 14 & 12.6 \\
 $\tau$~Sco     & 4.475 $\pm$ 0.073 & 4.31 $\pm$ 0.16 &  10$\pm$2  &  2 & 14.7 &  5 & 14.5 &  5 & 14.7 \\
 $\delta$~Sco   & 4.438 $\pm$ 0.033 & 4.58 $\pm$ 0.14 & 148$\pm$8  &  9 & 14.4 &  9 & 14.9 & 10 & 14.6 \\
 $\sigma$~Sco   & 4.443 $\pm$ 0.015 & 4.98 $\pm$ 0.12 &  56$\pm$15 &  8 & 17.7 &  8 & 18.0 & 10 & 17.2
\enddata
\tablecomments{References: (B94) \citet{1994A&AS..106..275B}, (E11) \citet{2011arXiv1110.5049E}.  
\vsini\, values adopted from \citet{1997A&A...319..811B} }
\end{deluxetable}

\clearpage
\begin{deluxetable}{llrrrrl}
\tabletypesize{\scriptsize}
\tablewidth{0pt}
\tablecaption{Stellar Parameters For A-type Upper Sco Members \label{tbl:atype_params}}
\tablehead{
  \colhead{HIP} &
  \colhead{Spectral} & 
  \colhead{A$_V$} & 
  \colhead{$\pi_{kin}$} &
  \colhead{log(\teff)} & 
  \colhead{log(L/\lsun)} &
  \colhead{Ref.} \\   
  \colhead{} & 
  \colhead{Type} & 
  \colhead{(mag)} & 
  \colhead{(mas)} & 
  \colhead{(dex)} & 
  \colhead{(dex)} &
  \colhead{}
}
\startdata
76310 & A0V      & 0.14 $\pm$ 0.03 & 8.22 $\pm$ 0.62 & 3.980 $\pm$ 0.031 & 1.271 $\pm$ 0.070 & 1 \\
77545 & A2/3V    & 1.25 $\pm$ 0.04 & 6.83 $\pm$ 0.62 & 3.937 $\pm$ 0.014 & 1.053 $\pm$ 0.081 & 2 \\
77815 & A5V      & 0.85 $\pm$ 0.07 & 6.89 $\pm$ 0.58 & 3.907 $\pm$ 0.007 & 1.195 $\pm$ 0.077 & 2 \\
77960 & A4IV/V   & 0.75 $\pm$ 0.02 & 8.56 $\pm$ 0.71 & 3.918 $\pm$ 0.012 & 0.973 $\pm$ 0.093 & 2 \\
78099 & A0V      & 0.59 $\pm$ 0.03 & 7.42 $\pm$ 0.59 & 3.980 $\pm$ 0.031 & 1.367 $\pm$ 0.073 & 2 \\
78196 & A0V      & 0.00 $\pm$ 0.04 & 8.45 $\pm$ 0.63 & 3.980 $\pm$ 0.031 & 1.302 $\pm$ 0.070 & 3 \\
78494 & A2mA7-F2 & 0.75 $\pm$ 0.06 & 7.26 $\pm$ 0.57 & 3.943 $\pm$ 0.016 & 1.395 $\pm$ 0.077 & 2 \\
78809 & A1Vnn    & 0.21 $\pm$ 0.05 & 7.29 $\pm$ 0.57 & 3.964 $\pm$ 0.019 & 1.244 $\pm$ 0.071 & 1 \\
\enddata
\tablecomments{Uncertainties in log(\teff) are determined assuming a spectral type uncertainty of 1 subtype.\\
Spectral Type References: 
(1) \citet{2001A&A...373..625P},
(2) \citet{1988mcts.book.....H},
(3) \citet{1982mcts.book.....H},
(4) \citet{2003AJ....126.2971V},
(5) \citet{1972AJ.....77..474G},
(6) \citet{1967ApJ...147.1003G}, 
(7) \citet{1981ApJS...45..437A},
(8) \citet{1998AJ....116.2530G} \\
$^*$ We list the hydrogen type in the table, the full type from \citet{1998AJ....116.2530G} is ``kA1 hA3 mA3 Vaer Bd1 $\le$ Nem1''
}
\end{deluxetable}

\begin{deluxetable}{lr}
\tabletypesize{\scriptsize}
\tablewidth{0pt}
\tablecaption{Age Constraints for Upper Sco from the A-type Main Sequence Turn-on\label{tbl:astar_ages}}
\tablehead{
  \colhead{Evolutionary} &
  \colhead{Age} \\
  \colhead{Tracks} & 
  \colhead{(Myr)} }
\startdata
 Dartmouth        & 10$\pm$2 \\
 Yonsei-Yale      & 10$\pm$2 \\
 SDF00            & 12$\pm$3 \\
 DM97             &  9$\pm$2 \\
\hline \\
 Median           & 10       \\
 Statistical Unc. &   $\pm$1 \\
 Systematic Unc.  &   $\pm$1
\enddata
\tablecomments{
The overall uncertainty is the 68\% C.L. of the median \citep{2001ApJ...549....1G} 
(statistical) and the dispersion in ages (systematic).}
\end{deluxetable}
\clearpage

\begin{landscape}
\begin{deluxetable}{lllrrrrrrrrr}
\tabletypesize{\scriptsize}
\tablewidth{0pt}
\tablecaption{Stellar Parameters for G-Type Upper Sco Members\label{tbl:pz99_gk}}
\tablehead{
  \colhead{Object} &
  \colhead{TYC} &
  \colhead{SpT} &
  \colhead{$\pi_{kin}$} &
  \colhead{A$_V$} & 
  \colhead{log(\teff)} & 
  \colhead{log(L/\lsun)} &
  \colhead{D08} & 
  \colhead{YY04} & 
  \colhead{S00} & 
  \colhead{DM97} &
  \colhead{Ref.} \\
  \colhead{} &
  \colhead{} &
  \colhead{} &
  \colhead{(mas)} &
  \colhead{(mag)} & 
  \colhead{(dex)} & 
  \colhead{(dex)} & 
  \colhead{(Myr)} & 
  \colhead{(Myr)} & 
  \colhead{(Myr)} & 
  \colhead{(Myr)} & 
  \colhead{} \\
}
\startdata
{[}PZ99]~J155548.7-251223 & 6783-2045-1 & G3      & 6.56 $\pm$ 0.85 & 0.70 $\pm$ 0.17 & 3.756 & 0.38 $\pm$ 0.13 & 11 & 12 & 18 & 12 & (1) \\
{[}PZ99]~J161618.0-233947 & 6793-1406-1 & G7      & 7.17 $\pm$ 0.81 & 0.73 $\pm$ 0.06 & 3.742 & 0.34 $\pm$ 0.10 &  9 & 10 & 16 &  9 & (1) \\
{[}PZ99]~J160843.4-260216 & 6784-39-1   & G7      & 7.39 $\pm$ 0.66 & 0.65 $\pm$ 0.09 & 3.742 & 0.36 $\pm$ 0.09 &  9 & 10 & 15 &  9 & (1) \\
{[}PZ99]~J161459.2-275023 & 6801-186-1  & G5      & 7.66 $\pm$ 0.92 & 0.73 $\pm$ 0.31 & 3.750 & 0.08 $\pm$ 0.16 & 23 & 23 & 30 & 21 & (1) \\
{[}PZ99]~J160040.6-220032 & 6212-1183-1 & G9      & 6.36 $\pm$ 1.06 & 0.65 $\pm$ 0.17 & 3.728 & 0.27 $\pm$ 0.16 &  8 &  9 & 15 &  7 & (1) \\
{[}PZ99]~J160000.7-250941 & 6783-1747-1 & G0      & 6.86 $\pm$ 0.65 & 0.36 $\pm$ 0.19 & 3.774 & 0.19 $\pm$ 0.11 & 24 & 24 & ... & 22 & (1) \\
{[}PZ99]~J160158.2-200811 & 6208-1543-1 & G5      & 6.31 $\pm$ 0.77 & 1.22 $\pm$ 0.25 & 3.750 & 0.65 $\pm$ 0.15 &  5 &  6 & 10 &  5 & (1) \\
{[}PZ99]~J155812.7-232835 & 6779-780-1  & G2      & 6.60 $\pm$ 0.77 & 0.86 $\pm$ 0.18 & 3.762 & 0.49 $\pm$ 0.12 & 10 & 10 & 16 & 10 & (1) \\
\enddata
\tablecomments{
Spectral Type References -- (1) \citet{1999AJ....117.2381P}, (2) \citet{2006A&A...460..695T}, (3) \citet{1982mcts.book.....H}, (4) \citet{1988mcts.book.....H} \\
}
\end{deluxetable}
\clearpage
\end{landscape}

\begin{deluxetable}{lrrr}
\tabletypesize{\scriptsize}
\tablewidth{0pt}
\tablecaption{Median Ages of the G-Type Upper Sco Members\label{tbl:us_gstar_ages}}
\tablehead{
  \colhead{Evolutionary} &
  \colhead{Age} \\
  \colhead{Tracks} & 
  \colhead{(Myr)} \\
}
\startdata
 Dartmouth   &  8$\pm$4 \\
 Yonsei-Yale &  9$\pm$3 \\
 SDF00       & 14$\pm$4 \\
 DM97        &  8$\pm$4 \\
\hline \\
Median           &       9 \\
Statistical Unc. &  $\pm$2 \\
Systematic Unc.  &  $\pm$3
\enddata
\tablecomments{
Uncertainties represent the 1$\sigma$ dispersion in the Chauvenet's-criterion clipped 
ages \citep{2003drea.book.....B}.  The uncertainty in the median is 
the 68\% confidence limit as described in \cite{2001ApJ...549....1G} (statistical) and 
the dispersion in ages (systematic).
}
\end{deluxetable}

\begin{deluxetable}{llrrr}
\tabletypesize{\scriptsize}
\tablewidth{0pt}
\tablecaption{Properties of Upper Sco Members Used For Expansion Age\label{tbl:us_expand}}
\tablehead{
  \colhead{HIP} &
  \colhead{d}  & 
  \colhead{v$_{rad}$(pred.)} &
  \colhead{v$_{rad}$(obs.)} &
  \colhead{Ref.} \\
  \colhead{} &
  \colhead{(pc)}  & 
  \colhead{(km~s$^{-1}$} & 
  \colhead{(km~s$^{-1}$} & 
  \colhead{} \\
}
\startdata
76071 & 175 $\pm$ 20 & -5.3 & -6.1 $\pm$ 2.3 & 1 \\
76310 & 151 $\pm$ 13 & -4.4 & -2.5 $\pm$ 0.6 & 1 \\
76503 & 190 $\pm$ 13 & -4.4 &  3.7 $\pm$ 20.0 & 2 \\
76633 & 145 $\pm$ 18 & -6.2 & -2.8 $\pm$ 2.2 & 3 \\
77635 & 152 $\pm$ 6  & -5.0 & -3.0 $\pm$ 4.7 & 3 \\
77813 & 105 $\pm$ 14 & -6.8 & -5.4 $\pm$ 0.4 & 4 \\
77840 & 154 $\pm$ 12 & -5.2 & -9.3 $\pm$ 1.6 & 3 \\
77858 & 129 $\pm$ 9  & -5.4 & -6.3 $\pm$ 20.0 & 2 \\
\enddata \\
\tablecomments{
Distances are derived using trigonometric parallaxes from \cite{2007A&A...474..653V}.
References: (1) \citet{2011arXiv1110.0536D}; (2) 
\citet{1995A&AS..114..269D}; (3) \citet{2006AstL...32..759G};
(4) \citet{2011ApJ...738..122C}; (5) \citet{2000A&AS..142..217B};
}
\end{deluxetable}

\begin{deluxetable}{lr}
\tabletypesize{\scriptsize}
\tablewidth{0pt}
\tablecaption{Upper Sco Age Estimates\label{tbl:us_ages}}
\tablehead{
  \colhead{Sample} & 
  \colhead{Age} \\
  \colhead{} & 
  \colhead{(Myr)} }
\startdata
 F-Type PMS                   & 13$\pm$1 \\
 Main-sequence Turn-off       & 10$\pm$2 \\
 Antares                      & 12$\pm$2 \\
 A-Type Turn-on               & 10$\pm$3 \\
 G-Type PMS                   &  9$\pm$2 \\
 \hline \\
 Adopted Age                  & 11       \\
 Statistical Unc.             & $\pm$1   \\
 Systematic  Unc.             & $\pm$2
\enddata
\end{deluxetable}

\begin{deluxetable}{lrr}
\tabletypesize{\scriptsize}
\tablewidth{0pt}
\tablecaption{Revised Mass Estimates for Substellar Objects in Upper Sco\label{tbl:substellar}}
\tablehead{
  \colhead{Object} &
  \colhead{Mass at 11~Myr} \\
  \colhead{} &
  \colhead{(M$_{\rm jup}$)} }
\startdata
 [PZ99]~J160930.3-210459B & 14$^{+2}_{-3}$ \\
 GSC~06214-00210B         & 17$\pm$3 \\
 HIP~78530B               & 30$^{+17}_{-8}$ \\
 UScoCTIO~108B            & 16$^{+3}_{-2}$ \\
 Oph~1622-2405A           & 53$^{+9}_{-7}$ \\
 Oph~1622-2405B           & 21$\pm$3
\enddata
\tablecomments{Masses estimated using the DUSTY models of 
\citet{2000ApJ...542..464C,2002A&A...382..563B} and an age for Upper Sco of 11$\pm$2~Myr.
}
\end{deluxetable}

\end{document}